\documentclass[letterpaper, 10 pt]{IEEEtran}
\IEEEoverridecommandlockouts

\usepackage{graphicx}
\usepackage{amsmath}
\usepackage{amssymb}
\usepackage{comment}
\usepackage{amsthm}
\usepackage{bbm}
\usepackage{here}
\usepackage{color}

\theoremstyle{definition}
\newtheorem{thm}{Theorem}
\newtheorem{dfn}[thm]{Definition}
\newtheorem{lem}[thm]{Lemma}
\newtheorem{prop}[thm]{Proposition}
\newtheorem{asm}[thm]{Assumption}

\newcommand{\av}{{\rm av}}
\newcommand{\intra}{{\rm intra}}
\newcommand{\inter}{{\rm inter}}
\newcommand{\FB}{{\rm FB}}
\newcommand{\PM}{{\rm PM}}
\newcommand{\argmax}{{\rm argmax}}

\newcommand{\T}{{\mathrm{T}}}
\newcommand{\tB}{\tilde{B}}

\newcommand{\diag}{\mathrm{diag}}
\newcommand{\blkdiag}{\text{blk-diag}}
\newcommand{\eigmax}{\lambda_{\rm max}}

\newcommand{\subjectto}{\mathrm{s.t.}}
\newcommand{\dist}{\mathrm{dist}}

\newcommand{\Graph}{\mathcal{G}}
\newcommand{\Tree}{\mathcal{T}}
\newcommand{\Nodes}{\mathcal{V}}
\newcommand{\Edges}{\mathcal{E}}
\newcommand{\Cluster}{\mathcal{C}}
\newcommand{\Real}{\mathbb{R}}
\newcommand{\Torus}{\mathbb{T}}
\newcommand{\CSM}{\mathcal{S}_\Pi} 
\newcommand{\ones}{\boldsymbol{1}}
\newcommand{\zeros}{\boldsymbol{0}}

\newcommand{\Set}[1]{\{#1\}}
\newcommand{\Vector}[2]{[#1\,\cdots\,#2]}
\newcommand{\Range}[2]{#1,\,\dots\,,#2}
\newcommand{\Comp}[2]{#1 \mid #2}

\title{\LARGE \bf
 Cluster Synchronization and Phase Cohesiveness of Kuramoto Oscillators via Mean-phase Feedback Control and Pacemakers
}

\author{Ryota Kokubo, Rui Kato, and Hideaki Ishii
\thanks{This work was supported in part by JSPS under Grant-in-Aid for Scientific Research Grant No.~22H01508 and 24K00844, and by JST under ASPIRE Grant No.~JPMJAP2402.}
\thanks{R. Kokubo is with the Department of Computer Science, Tokyo Institute of Technology, Yokohama 226-8502, Japan. (email: \tt \small kokubo.r.ab@m.titech.ac.jp)}
\thanks{R. Kato is with the Laboratoire d'Automatique, de G\'enie des Proc\'ed\'es et de G\'enie Pharmaceutique (LAGEPP), Universit\'e Claude Bernard Lyon 1, Villeurbanne 69622, France. (email: \tt \small rui.kato@univ-lyon1.fr)}
\thanks{H. Ishii is with the Department of Information Physics and Computing, The University of Tokyo, Tokyo 113-8656, Japan. (email: \tt \small hideaki\_ishii@ipc.i.u-tokyo.ac.jp)}}

\begin{document}
\maketitle
\thispagestyle{empty}

\begin{abstract}
    Brain networks typically exhibit characteristic synchronization patterns where several synchronized clusters coexist.
    On the other hand, neurological disorders are considered to be related to pathological synchronization such as excessive synchronization of large populations of neurons.
    Motivated by these phenomena, this paper presents two approaches to control the cluster synchronization and the cluster phase cohesiveness of Kuramoto oscillators.
    One is based on feeding back the mean phases to the clusters, and the other is based on the use of pacemakers.
    First, we show conditions on the feedback gains and the pacemaker weights for the network to achieve cluster synchronization.
    Then, we propose a method to find optimal feedback gains through convex optimization.
    Second, we show conditions on the feedback gains and the pacemaker weights for the network to achieve cluster phase cohesiveness.
    A numerical example demonstrates the effectiveness of the proposed methods.
\end{abstract}


\section{Introduction}
\label{sec:introduction}
Synchronization phenomena in networked systems can be found and employed in various fields.
In biological and medical domains, synchronic flashing of firefly flocks \cite{moiseff-2000} and
synchronization of heartbeat and respiratory rhythms \cite{schafer-1998} are observed.
Examples in engineering domains include clock synchronization in wireless sensor networks \cite{simeone-2008, iori-2020}
and phase synchronization in smart grids \cite{dorfler-2013}.
Recently, synchronization phenomena in brain networks have garnered significant attention, owing to 
advances in measurement technologies such as functional magnetic resonance imaging (fMRI) \cite{bassett-2017}.
Particularly noteworthy is the concept of \emph{cluster synchronization}, where neural activities are synchronized within certain regions of a healthy brain while remaining unsynchronized in others \cite{gray-1994}.
Conversely, brain networks affected by conditions such as epilepsy or Parkinson's disease exhibit excessive synchronization among regions that are typically unsynchronized \cite{lehnertz-2009,hammond-2007}.
Hence, exploring control methods to stabilize cluster synchronization holds promise for advancing treatments for such neurological diseases.

In this paper, we consider a networked system based on Kuramoto oscillators \cite{kuramoto-1975}.
Originally proposed as a model of chemically coupled oscillators, Kuramoto oscillators have found 
widespread utility as a model for various synchronization phenomena.
For instance, by representing the electrical activity of neurons as time-delayed Kuramoto oscillators, 
it is possible to replicate the correlation of neural activities 
between different brain network regions \cite{cabral-2011}.
For this class of systems, there are studies related to cluster synchronization. 
For example, the work \cite{menara-cns2020} derived a sufficient condition for exponential stability of the cluster synchronization manifold, by analyzing the influence from different clusters as perturbations.
In \cite{menara-acc2019}, a condition for approximate cluster synchronization was obtained using a method based on the small gain theorem.
Furthermore, in \cite{kato-2021}, sufficient conditions for the exponential stability of cluster synchronization using an averaging method were studied.
In a similar setting, the authors of \cite{cao-2021} investigated partial synchronization using several cohesiveness measures.
It is also known that in networked nonlinear systems, the presence of time delays in the couplings can induce cluster synchronization (e.g., \cite{ryono-2015}) and its approximate counterpart \cite{su-2019}.

On the other hand, several works focused on stabilizing Kuramoto oscillator networks through different control approaches.
The work \cite{doyle-2013} proposed a method to stabilize the full synchronization by inputting a signal of a pacemaker that continues to oscillate at a constant frequency to oscillators.
This method is based on feedforward control and requires prior knowledge of each node's natural frequency.
For cluster synchronization, the method proposed in \cite{menara-cdc2019} is to adjust system parameters such as edge weights and node natural frequencies so as to satisfy the approximate cluster synchronization condition in \cite{menara-acc2019}; it determines parameter corrections through an optimization problem to minimize the number of corrections.
Moreover, in \cite{qin-2022}, vibrational control is applied to edge weights for cluster synchronization, and its relation with deep brain stimulation was studied.

In this paper, we propose two approaches for stabilizing the cluster synchronization state in Kuramoto oscillators by introducing external inputs to the clusters.
One is based on the feedback of the mean of the phases within each cluster.
While measurements of the phases are needed, we develop a convex optimization method for finding the feedback gains to minimize the number of nodes to which control should be applied.
The other approach is based on pacemakers as in \cite{doyle-2013}, where for each cluster, inputs are generated by a pacemaker oscillating at the same frequency as the nodes' natural frequencies in the cluster.
This is a feedforward approach, and we show that stabilization can be achieved if the pacemaker weights are sufficiently large.

The two control methods proposed in this paper, based on pacemakers and mean-phase feedback control, are conceptually related to recent developments in brain stimulation techniques. A representative example of such external input is deep brain stimulation (DBS), where periodic signals are applied to specific regions of the brain to modulate abnormal activities. DBS has been widely adopted in the clinical treatment of neurological disorders such as Parkinson’s disease, and is typically implemented as an open-loop system, delivering constant-frequency signals without using real-time measurements. The work \cite{carron-2013} provides a comprehensive overview of such approaches, discussing their mechanisms and effects based on both experimental observations and theoretical models. While open-loop methods like DBS lack adaptability to changing internal states, their simplicity and demonstrated effectiveness support the feasibility of pacemaker-type external inputs as a practical means of regulating network synchronization.

In contrast, the mean-phase feedback control falls into the broader class of closed-loop approaches, where the input signal is updated based on the current state of the system. Several studies have shown that such feedback-based approaches can improve the efficiency and adaptability of stimulation. For example, \cite{little-2013} proposed a method for adjusting the input amplitude in real time based on measurable signals related to brain activities, reporting improved performance over conventional DBS. In \cite{hebb-2014}, various design methods for closed-loop stimulation were reviewed, including proportional and optimization-based control methods. Furthermore, \cite{mirza-2019} discussed the possibility of using chemical sensing as a feedback mechanism, highlighting the potential of combining sensing technologies with control algorithms.

These studies indicate that both open-loop and closed-loop approaches, similar to those proposed in this study, are actively being investigated in applied contexts. This supports the relevance of exploring such control methods not only from a theoretical perspective but also in view of their applicability to real-world systems.

The rest of the paper is organized as follows.
In Section~\ref{sec:preliminaries}, we describe the problem setting.
The main results on cluster synchronization via mean-phase feedback control and pacemakers are provided in Sections~\ref{sec:feedback}~and~\ref{sec:pacemaker}, respectively.
Further results on cluster phase cohesiveness via the two control approaches are provided in Section~\ref{sec:cluster_phase_cohesiveness}.
Our results are demonstrated by a numerical example in Section~\ref{sec:numerical example}.
Finally, we conclude the paper in Section~\ref{sec:conclusion}.
This paper is based on the preliminary version \cite{kokubo-2024} and contains the full proofs and additional results.

\textit{Notations}: Let $\mathbb{R}$ and $\mathbb{Z}$ be the sets of real numbers and integers, respectively.
Let $\mathbb{S}^1$ be the unit circle and $\mathbb{T}^n = \mathbb{S}^1 \times \cdots\times\mathbb{S}^1$ the $n$-torus.
We identify $\mathbb{S}^1$ as the group $\mathbb{R}/(2\pi\mathbb{Z})$ with addition modulo $2\pi$, that is, $\theta\in\mathbb{S}^1$ implies $\theta + 2\pi k \in\mathbb{S}^1$ for any $k\in\mathbb{Z}$.
With this notation, for $\theta,\,\tilde{\theta} \in \mathbb{S}^1$, their difference $\theta - \tilde{\theta}$ again belongs to $\mathbb{S}^1$.
The $n$-vector whose entries are all $1$ is denoted by $\ones_n$.
We define the distance from a point $x\in\mathbb{R}^n$ to a subset $\mathcal{S}\subset\mathbb{R}^n$ by $\mathrm{dist}(x, \mathcal{S}) := \inf_{y\in S}\|x-y\|$.
For a matrix $A$, let $A^+$ be its Moore--Penrose inverse.
For a symmetric matrix $A$, let $\lambda_{\max}(A)$ be the maximum eigenvalue of $A$.

\section{Preliminaries}
\label{sec:preliminaries}
In this section, we introduce the Kuramoto oscillators and the notion of cluster synchronization.
Then, we characterize the stability condition, which will be used in our analysis.

\subsection{Problem Setup}
We consider the so-called Kuramoto oscillators for the coupled oscillators consisting of $n$ nodes.
The connectivity structure among the oscillators is represented by an undirected graph $\Graph := (\Nodes, \Edges)$, 
where the node set $\Nodes := \Set{\Range{1}{n}}$ is the set of oscillators 
and the edge set $\Edges \subseteq \Nodes\times\Nodes$ represents the coupling structure of oscillators.
We assume that the overall graph $\Graph$ is connected.
Let $A := [a_{ij}] \in \Real^{n \times n}$ be the weighted adjacency matrix of $\Graph$, 
where $a_{ij} > 0$ if $(i,j)\in\Edges$ and $a_{ij} = 0$ otherwise.
Denote the phase of the oscillator $i$ at time $t \geq 0$ by $\theta_i(t) \in\mathbb{S}^1$ and 
the natural frequency of the oscillator $i$ by $\omega_i \in \Real$. 
The dynamics of the phase $\theta_i(t)$ is given by
\begin{equation}
\label{eq:original_Kuramoto_Oscillators}
    \dot{\theta}_i(t) = \omega_i + \sum_{j=1}^na_{ij}\sin(\theta_j(t)-\theta_i(t)),\:\:\:i\in\Nodes.
\end{equation}

These Kuramoto oscillators are said to be in full synchronization when all phases are equal, i.e. $\theta_i(t) = \theta_j(t)$ for $i,j\in\Nodes$.
To achieve such synchronization, various stability conditions have been studied \cite{bullo-lns}.
In particular, a necessary condition to achieve full synchronization is that all natural frequencies are identical, i.e., $\omega_i = \omega_j$ for $i,j\in\mathcal{V}$.

In this paper, our focus is to realize cluster synchronization, where the nodes form groups and only within the individual groups, the nodes synchronize with each other.
To this end, we introduce several additional notions for the problem formulation.
First, consider a partition $\Pi:=\Set{\Range{\Cluster_1}{\Cluster_m}}$ of the node set $\Nodes$,
where $\Cluster_k \subset \Nodes$, $\bigcup_{k=1}^m \Cluster_k = \Nodes$, and 
$\Cluster_k \cap \Cluster_l = \emptyset$ if $k \neq l$. Each subset $\Cluster_k$ is called a cluster, and the induced subgraph of each cluster is assumed to be connected.
The clusters remain invariant over time in this paper.
Hence, for ease of notation, we reorder the indices of the nodes so that those in $\Cluster_1$ are indexed as $1,\,2,\,\dots\,,|\Cluster_1|$, and those in $\Cluster_2$ are $|\Cluster_1|+1,\,|\Cluster_1|+2,\,\dots\,,|\Cluster_1|+|\Cluster_2|$, and so on.

Now, the cluster synchronization manifold $\CSM$ with respect to the partition $\Pi$ is defined by
\begin{equation}
    \label{eq:cluster_synchronization_manifold}
    \CSM := \Set{\Comp{\theta \in \Torus^n}{\theta_i=\theta_j,\,i,j\in\Cluster_k,\,k\in\Set{\Range{1}{m}}}}.
\end{equation}
It is known that the cluster synchronization manifold is an invariant manifold if and only if the two conditions in the following assumption are satisfied\footnote{The 'only if' part is true whenever the natural frequencies are sufficiently different among clusters.} \cite{tiberi-cdc2017}:
\begin{asm}
    \label{asm:cluster_sync_asm}
    \begin{itemize}
        \item[(i)] $\omega_i = \Omega_k$ for all $i\in\Cluster_k$ and all $ k\in\Set{\Range{1}{m}}$, where $\Omega_k \in \Real$ denotes the natural frequency of cluster $\Cluster_k$.
        \item[(ii)] $\sum_{h\in\Cluster_l}a_{ih} = \sum_{h\in\Cluster_l}a_{jh}$
        for all $i,j\in\Cluster_k$ and all $k,l\in\Set{\Range{1}{m}}$ with $k\neq l$.
    \end{itemize}
\end{asm}
In Assumption~\ref{asm:cluster_sync_asm}, condition (i) indicates that the natural frequencies of oscillators in the same cluster are equal.
On the other hand, condition (ii) is known as the external equitable partition (EEP) condition \cite{schaub-2016} and indicates that in each cluster, the sum of edge weights from neighboring oscillators of different clusters is equal among oscillators in the same cluster.
In Sections \ref{sec:feedback} and \ref{sec:pacemaker}, we assume that these two conditions hold.
In Section V, we will study the more relaxed version of the problem called cluster phase cohesiveness, where within each cluster, phases must remain close to each other with some error bound; in this case, the two conditions in Assumption \ref{asm:cluster_sync_asm} need not hold strictly.

We rewrite \eqref{eq:original_Kuramoto_Oscillators} in vector form.
Let the phase vector be $\theta(t) := [\theta_1(t) \cdots \theta_n(t)]^\T = [\theta^{(1)}(t)^\T \cdots \theta^{(m)}(t)^\T]^\T$, where $\theta^{(k)}(t)$ is the phase vector of cluster $\Cluster_k$ for $k\in\{1,\,\dots\,,m\}$. After choosing an enumeration and an orientation of each edge, let $B \in \Real^{n\times|\Edges|}$ be an oriented incidence matrix of $\Graph$ \cite{bullo-lns}.
We further define the diagonal weight matrix $W:= \diag(\Set{a_{ij}}_{(i,j)\in\Edges})\in \Real^{|\Edges|\times|\Edges|}$.
Then, \eqref{eq:original_Kuramoto_Oscillators} can be described as
\begin{equation}
\label{eq:original_Kuramoto_Oscillators_vectorform}
    \dot{\theta}(t) = \omega - BW\sin(B^{\mathrm{T}}\theta(t)), 
\end{equation}
where $\omega:=\Vector{\omega_1}{\omega_n}^\T$ is the natural frequency vector.

\subsection{Reformulation to Phase Difference Dynamics}
In this subsection, we transform the phase dynamics \eqref{eq:original_Kuramoto_Oscillators_vectorform} into the dynamics of the phase differences among the oscillators by taking account of the cluster structure \cite{menara-cns2020, kato-2021}.

\begin{figure}[tbp]
\centering
\begin{minipage}{0.4\hsize}
    \vspace{4mm}
    \centering
    \includegraphics[width=0.95\hsize]{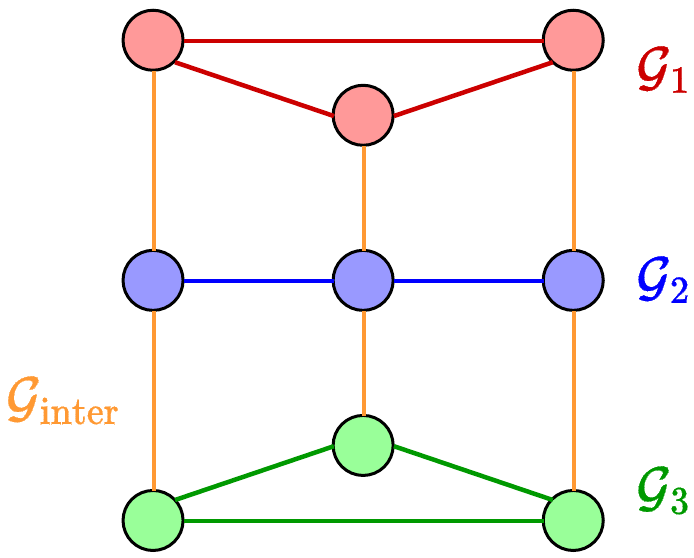}
    \caption{Graph $\Graph$ and its subgraphs}
    \label{fig:original_cluster_division}
\end{minipage}
\hspace{6mm}
\begin{minipage}{0.4\hsize}
    \vspace{4mm}
    \centering
    \includegraphics[width=0.95\hsize]{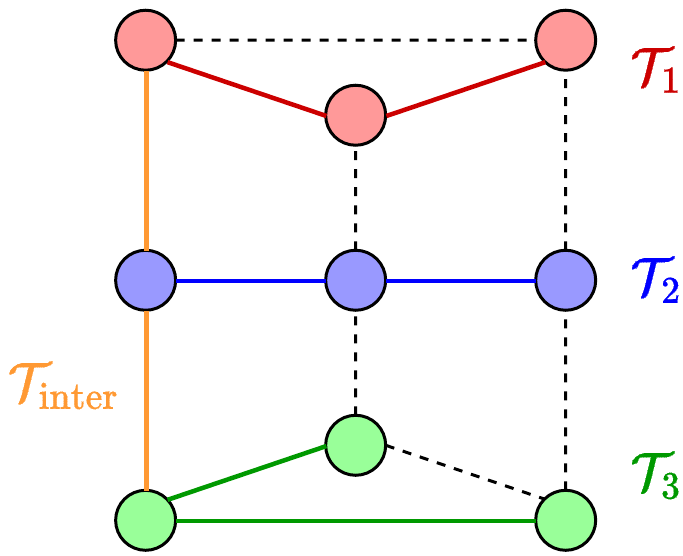}
    \caption{Spanning tree $\Tree$ and its decomposition}
    \label{fig:original_spanning_tree}
\end{minipage}
\vspace{-2mm}
\end{figure}

First, we introduce the induced subgraph $\Graph_k$ corresponding to cluster $\Cluster_k$ and the subgraph $\Graph_\inter$ corresponding to the connections among clusters.
Then, we can partition $\Graph$ as
\begin{equation}
\label{eq:original_graph_division}
    \Graph = \bigcup_{k=1}^m \Graph_k \cup \Graph_\inter,
\end{equation}
where $\Graph_k := (\Cluster_k,\Edges_k)$ with $\Edges_k := \Set{\Comp{(i,j)\in\Edges}{i,j\in\Cluster_k}}$ and
$\Graph_\inter := (\Nodes, \Edges_\inter)$ with $\Edges_\inter := \Edges\setminus\bigcup_{k=1}^m\Edges_k$.
We now express the oriented incidence matrix $B$ and the diagonal weight matrix $W$ of the graph $\Graph$ using the oriented incidence matrices and the diagonal weight matrices of the subgraphs $\Graph_k$ and $\Graph_\inter$ denoted by $B_k$, $B_\inter$, $W_k$, and $W_\inter$, respectively.
First, let $B_\intra := \blkdiag(\Range{B_1}{B_m})$. Then, we obtain
\(
    B = \begin{bmatrix}B_\intra & B_\inter\end{bmatrix}
\)
and 
\(
    W = \blkdiag(\Range{W_1}{W_m},\,W_\inter)
\).
Fig.~\ref{fig:original_cluster_division} shows an example of partitioning a graph into subgraphs.
Here, the nodes with the same color indicate that they are in the same cluster.

Next, we introduce an alternative expression for the graphs $\Graph$ and $\Graph_k$ by defining their corresponding spanning trees as $\Tree:=(\Nodes,\tilde{\Edges})$ and $\Tree_k := (\Cluster_k,\tilde{\Edges}_k)$, respectively, 
where the edge set $\tilde{\Edges}$ contains edges from all $\tilde{\Edges}_k$.
Let $\Tree_\inter := (\Nodes,\tilde{\Edges}_\inter)$ with 
$\tilde{\Edges}_\inter := \tilde{\Edges}\setminus\bigcup_{k=1}^m\tilde{\Edges}_k$.
For the graph of Fig.~\ref{fig:original_cluster_division}, an example of configuring $\Tree_k$ and $\Tree_\inter$ in this way is shown.
Similarly to \eqref{eq:original_graph_division}, we can express the spanning tree $\Tree$ as $\Tree = \bigcup_{k=1}^m\Tree_k \cup \Tree_\inter$.

Based on the enumeration and the orientation of each edge, let $\tB \in \Real^{n\times(n-1)}$ be an oriented incidence matrix of $\Tree$.
The oriented incidence matrices of $\Tree_k$ and $\Tree_\inter$ corresponding to $\tB$ are denoted as $\tB_k$ and $\tB_\inter$, respectively.
Hence, let $\tB_\intra := \blkdiag(\Range{\tB_1}{\tB_m})$, and then we can decompose $\tilde{B}$ in the form
\(
    \tB = \begin{bmatrix} \tB_\intra & \tB_\inter \end{bmatrix}
\).
Due to properties of spanning trees, there exists a matrix $P$ such that $B^\T = P\tB^\T$ and 
there exists a matrix $Q_k$ such that $B_k^\T = Q_k\tB_k^\T$.
Now, the relation can be expressed in the block matrix form as
\begin{equation*}
    \begin{bmatrix}
        B_\intra^\T \vspace{0.1em}\\ B_\inter^\T
    \end{bmatrix}
    =
    \begin{bmatrix}
        P_1 & 0 \\ P_2 & P_3
    \end{bmatrix}
    \begin{bmatrix}
        \tB_\intra^\T \\ \tB_\inter^\T
    \end{bmatrix}.
\end{equation*}
In particular, we note that $P_1 = \blkdiag(\Range{Q_1}{Q_m})$.

Next, let $x^{(k)}(t)$ be the phase difference vector associated with $\Tree_k$, and $z(t)$ be the phase difference vector associated with $\Tree_\inter$.
They can be obtained as
\begin{equation}
\label{eq:def_x}
x^{(k)}(t) := \tB_k^\T\theta^{(k)}(t),\,z(t) := \tB_\inter^\T\theta(t).
\end{equation}
Also, let $x(t) := \Vector{x^{(1)}(t)^\T}{x^{(m)}(t)^\T}^\T.$

To write the system \eqref{eq:original_Kuramoto_Oscillators_vectorform} using these new states, we split  \eqref{eq:original_Kuramoto_Oscillators_vectorform} into phase vectors $\theta^{(k)}$ of each cluster as
\begin{equation}
\label{eq:original_phase_dynamics}
    \begin{bmatrix}
        \dot{\theta}^{(1)}(t) \\ \vdots \\ \dot{\theta}^{(m)}(t)
    \end{bmatrix}
    = \omega + 
    \begin{bmatrix}
        F_{\intra,\theta}^{(1)}(\theta^{(1)}(t)) \\ \vdots \\ F_{\intra,\theta}^{(m)}(\theta^{(m)}(t))
    \end{bmatrix}
    + F_{\inter,\theta}(\theta(t)),
\end{equation}
where 
\begin{align*}
    F_{\intra,\theta}^{(k)}(\theta^{(k)}(t)) &:= -B_kW_k\sin(B_k^\T\theta^{(k)}(t)),\\
    F_{\inter,\theta}(\theta(t)) &:= -B_\inter W_\inter\sin(B_\inter^\T\theta(t)).
\end{align*}
Multiplying \eqref{eq:original_phase_dynamics} by $\tB_\intra$ from the left, we can derive the phase difference dynamics
\begin{equation}
\label{eq:original_phase_difference_dynamics}
    \begin{bmatrix}
        \dot{x}^{(1)}(t) \\ \vdots \\ \dot{x}^{(m)}(t)
    \end{bmatrix}
    =  
    \begin{bmatrix}
        F_{\intra,x}^{(1)}(x^{(1)}(t)) \\ \vdots \\ F_{\intra,x}^{(m)}(x^{(m)}(t))
    \end{bmatrix}
    + F_{\inter,x}(x(t), z(t)),
\end{equation}
where
\begin{align*}
    F_{\intra,x}^{(k)}(x^{(k)}(t)) := &-\tB_k^\T B_kW_k\sin(Q_kx^{(k)}(t)),\\
    F_{\inter,x}(x(t), z(t)) := &-\tB_\intra^\T B_\inter W_\inter\\
    &\times \sin(P_2x(t) + P_3z(t)).
\end{align*}
To obtain the above, we used the relations
\begin{align*}
    B_k^\T\theta^{(k)}(t) &= Q_k\tB_k^\T\theta^{(k)}(t) = Q_kx^{(k)}(t),\\
    B_\inter^\T\theta(t) &= (P_2\tB_\intra^\T+P_3\tB_\inter^\T)\theta(t)
    = P_2x(t)+P_3z(t),
\end{align*}
and the fact that $\tB_\intra^\T\omega = 0$ under Assumption~\ref{asm:cluster_sync_asm}~(i).
The first term on the right-hand side of \eqref{eq:original_phase_difference_dynamics} represents the effects from oscillators in the same cluster, and the second term represents the effects from oscillators in different clusters.

So far, we obtained the dynamics \eqref{eq:original_phase_difference_dynamics} of the phase differences for the uncontrolled case of the oscillator network.
In Sections III and IV, we will turn our attention to the controlled cases and derive similar equations for the dynamics.

\subsection{Stability of the Cluster Synchronization Manifold}
In this subsection, we define local exponential stability of the cluster synchronization manifold \eqref{eq:cluster_synchronization_manifold}.
Then, we provide a sufficient condition that guarantees the Kuramoto oscillators to achieve such stability.
\begin{dfn}
    The cluster synchronization manifold $\CSM$ of Kuramoto oscillators \eqref{eq:original_Kuramoto_Oscillators_vectorform} 
    is said to be locally exponentially stable if there exist $C \geq 1$, $\lambda > 0$, and $\delta > 0$ such that if 
    $\dist(\theta(0),\CSM) < \delta$, 
    then $\dist(\theta(t),\CSM) \leq C \mathrm{e}^{-\lambda t}\dist(\theta(0),\CSM)$ 
    for any $t \geq 0$.
\end{dfn}
It is important to note that the stability of the cluster synchronization manifold is equivalent to the stability of the equilibrium $x = 0$ of \eqref{eq:original_phase_difference_dynamics}.
Now, let $J_{\intra, k}$ be the Jacobian of $F_{\intra,x}^{(k)}(x^{(k)})$ in \eqref{eq:original_phase_dynamics} at $x^{(k)} = 0$.
It is known that under Assumption~\ref{asm:cluster_sync_asm}, $J_{\intra,k}$ is a Hurwitz matrix \cite{menara-cns2020}.
Hence, let $X_k \succ 0$ be the solution of the Lyapunov equation 
\begin{equation}
    \label{eq:lyapunov_eq}
    J_{\intra,k}^\T X_k + X_kJ_{\intra,k} = -I.
\end{equation}
Moreover, we define
$\gamma^{(kl)} := \kappa \tilde{\gamma}^{(kl)}$ 
for $k,l\in\Set{\Range{1}{m}}$, where $\kappa := \max_{r\in\Set{\Range{1}{m}}} 2(|\Cluster_r|-1)$ and 
\begin{equation}
    \label{eq:gamma_tilde}
\tilde{\gamma}^{(kl)} := \left\{\begin{array}{ll}
    \sum_{h=1,h\neq k}^m\sum_{j\in\Cluster_h} a_{ij} & \mathrm{if}\;\;k = l,\\
    \sum_{j\in\Cluster_l} a_{ij} & \mathrm{if}\;\;k \neq l,
\end{array}\right.
\end{equation}
with $i\in\Cluster_k$. Note that from Assumption~\ref{asm:cluster_sync_asm}~(ii), $\tilde{\gamma}^{(kl)}$ is nonnegative and uniquely determined regardless of $i$ in \eqref{eq:gamma_tilde}.

With these, define the matrix
\begin{equation}
\label{eq:original_S_matrix}
    S = [s_{kl}] := \left\{
    \begin{array}{ll}
        \left(\eigmax(X_k)\right)^{-1} - \gamma^{(kk)} &  \mathrm{if}\;\;k = l,\\
        -\gamma^{(kl)} & \mathrm{if}\;\;k \neq l.
    \end{array}
    \right.
\end{equation}
Based on this matrix $S$, a stability condition for the cluster synchronization manifold of Kuramoto oscillators without control inputs is given below \cite{menara-cns2020} (see also \cite{khalil-nonlinear}).
Recall that a square matrix is an $M$-matrix if all off-diagonal entries are nonpositive and the real parts of all eigenvalues are positive.
\begin{lem}[\cite{menara-cns2020}]
\label{lem:original_stability}
    Under Assumption~\ref{asm:cluster_sync_asm}, the cluster synchronization manifold is locally exponentially stable if the matrix $S$ 
    is an $M$-matrix.
\end{lem}
From the lemma above, we obtain the following proposition, which provides a sufficient condition for the matrix $S$ to be an $M$-matrix.
This means that it is more conservative, but it has an advantage for our purpose to design control signals for achieving cluster synchronization in the following sections.
In particular, the condition can be checked for each cluster, and hence we can find appropriate controllers for the clusters.
\begin{prop}
\label{prop:clusterwise_stability}
    Under Assumption~\ref{asm:cluster_sync_asm}, for the Kuramoto oscillators \eqref{eq:original_Kuramoto_Oscillators}, if $\eigmax(J_{\intra,k} + J_{\intra,k}^\T) < -2\gamma^{(kk)}$ for all $k\in\Set{\Range{1}{m}}$, then the cluster synchronization manifold is locally exponentially stable.
\end{prop}
\begin{proof}
    Based on Lemma \ref{lem:original_stability}, we must show that the matrix $S$ is an $M$-matrix.
    First, since $\gamma^{(kl)}$ are nonnegative, all off-diagonal entries of $S$ are nonpositive.
    Thus, it remains to show that the real parts of all eigenvalues of $S$ are positive.

    Because of the connectivity of $\Graph$, $\gamma^{(kk)} > 0$.
    Hence, $\eigmax(J_{\intra,k} + J_{\intra,k}^\T) < -2\gamma^{(kk)}$ implies $\eigmax(J_{\intra,k} + J_{\intra,k}^\T) < 0$ for any $k\in\Set{\Range{1}{m}}$.
    Then, from Theorem~2.4 in \cite{gajic-lyap1995} and \eqref{eq:lyapunov_eq}, the following holds:
    \begin{align}
    \nonumber
        \eigmax(X_k) &\leq \frac{1}{|\eigmax(J_{\intra,k}+J_{\intra,k}^\T)|}\\
    \label{eq:original_upperbound}
        &= \frac{1}{-\eigmax(J_{\intra,k}+J_{\intra,k}^\T)} < \frac{1}{2\gamma^{(kk)}}.
    \end{align}
    From the definition of $S$ in \eqref{eq:original_S_matrix}, we have
        \begin{equation*}
        \sum_{l=1,l\neq k}^m |s_{kl}| = 
        \sum_{l=1,l\neq k}^m \gamma^{(kl)} = 
        \sum_{l=1,l\neq k}^m \sum_{j\in\Cluster_l} a_{ij} = \gamma^{(kk)}.
    \end{equation*}
    Thus \eqref{eq:original_S_matrix} and \eqref{eq:original_upperbound} implies
    \begin{equation*}
        \: s_{kk} - \sum_{l=1,l\neq k}^m |s_{kl}| =(\eigmax(X_k))^{-1} - 2\gamma^{(kk)}> 0.
    \end{equation*}
    Now, for any eigenvalue $\lambda$ of the matrix $S$, from Gershgorin's circle theorem \cite{bullo-lns}, it holds
    \begin{equation*}
        \Re(\lambda) \geq \min_{k\in\Set{\Range{1}{m}}} s_{kk} - \sum_{l=1,l\neq k}^m |s_{kl}| > 0. \qedhere
    \end{equation*}
\end{proof}

Lemma~\ref{lem:original_stability} and Proposition~\ref{prop:clusterwise_stability} provide conditions under which the Kuramoto oscillators in (1) achieve cluster synchronization. 
In the literature, other conditions can be found as well \cite{menara-acc2019, kato-2021}.
In the following sections, however, we assume that such conditions do not hold and thus in the oscillator system, the clusters cannot synchronize, and to do so, it requires additional control mechanisms.
To this end, we introduce two approaches, based on mean-phase feedback control in Section III and pacemakers in Section IV.

\section{Stabilization by Mean-phase Feedback Control}
\label{sec:feedback}
In this section, we consider stabilizing the cluster synchronization manifold based on mean-phase feedback control.
This method is feedback-based in that the average of the phases within each cluster must be first computed in real-time 
and then used as an input to the nodes.
We will show that under this approach, it may be enough to apply it only at some nodes.

\subsection{Mean-phase Feedback Control}
For $k \in \Set{\Range{1}{m}}$, we define the mean phase of oscillators in the cluster $\Cluster_k$ by
\[
    \theta_{\av, k}(t) := \frac{1}{|\Cluster_k|}\sum_{i\in\Cluster_k}\theta_i(t).
\]
When the mean-phase feedback is introduced, the phase $\theta_i(t)$ of node $i$ is governed by
\begin{align}
\nonumber
    \dot{\theta_i}(t) =\; &\Omega_k + \sum_{j=1}^na_{ij}\sin(\theta_j(t) - \theta_i(t)) \\
\label{eq:feedback_Kuramoto_Oscillators}
    &+ g_i\sin(\theta_{\av, k}(t) - \theta_i(t)),\:\:i\in\Cluster_k,
\end{align}
where $g_i \geq 0$ is the feedback gain.
This control has an effect to strengthen the interactions among nodes in each cluster as it uses the mean phase of all nodes. 
That is, through this control, we can effectively increase the number and/or the weights of the edges connecting nodes in a cluster.
Such interactions help in forming local synchronization within each cluster, and increasing the gain $g_i$ will strengthen this effect.

We now rewrite the network in vector form.
It follows that for each cluster $k\in\Set{\Range{1}{m}}$, the phase error from the mean-phase within the cluster is given by
\begin{align*}
    &\;\theta_\av^{(k)}(t)\ones_{|\Cluster_k|} - \theta^{(k)}(t) = \frac{1}{|\Cluster_k|}\ones_{|\Cluster_k|}\ones_{|\Cluster_k|}^\T\theta^{(k)}(t) - \theta^{(k)}(t)\\
    &\text{\hspace{2cm}} = \frac{1}{|\Cluster_k|}\left(\ones_{|\Cluster_k|}\ones_{|\Cluster_k|}^\T - |\Cluster_k|I_{|\Cluster_k|}\right)\theta^{(k)}(t).
\end{align*}
Note that the matrix $\ones_{|\Cluster_k|}\ones_{|\Cluster_k|}^\T - |\Cluster_k|I_{|\Cluster_k|}$ on the far right-hand side is equal to the Laplacian matrix of the complete graph of order $|\Cluster_k|$; denote this Laplacian by $L_{\FB, k}$.
Hence, defining the matrix $G := \diag(\Range{g_1}{g_n})$ and the mean-phase vector $\theta_\av(t):=\Vector{\theta_\av^{(1)}(t)\ones_{|\Cluster_1|}^\T}{\theta_\av^{(m)}(t)\ones_{|\Cluster_m|}^\T}^\T$,
we can rewrite \eqref{eq:feedback_Kuramoto_Oscillators} in the vector form as
\[
    \dot{\theta}(t) = \omega - BW\sin(B^{\mathrm{T}}\theta(t)) - G\sin(L_\FB\theta(t)),
\]
where $L_\FB := \blkdiag\left(\Range{\frac{1}{|\Cluster_1|}L_{\FB,1}}{\frac{1}{|\Cluster_m|}L_{\FB,m}}\right)$.

The cluster-wise dynamics can be described by
\begin{align}
\nonumber
    \begin{bmatrix}
        \dot{\theta}^{(1)}(t) \\ \vdots \\ \dot{\theta}^{(m)}(t)
    \end{bmatrix}
    =\; &\omega + 
    \begin{bmatrix}
        F_{\intra,\theta}^{(1)}(\theta^{(1)}(t)) \\ \vdots \\ F_{\intra,\theta}^{(m)}(\theta^{(m)}(t))
    \end{bmatrix}
    + F_{\inter,\theta}(\theta(t))\\ 
\label{eq:feedback_phase_dynamics}
    &+
    \begin{bmatrix}
        F_{\FB,\theta}^{(1)}(\theta^{(1)}(t)) \\
        \vdots\\
        F_{\FB,\theta}^{(m)}(\theta^{(m)}(t))
    \end{bmatrix},
\end{align}
where 
\(
    F_{\FB,\theta}^{(k)}(\theta^{(k)}(t)) := G_k\sin\left(-\frac{1}{|\Cluster_k|}L_{\FB,k}\theta^{(k)}(t)\right),
\)
and $G_k$ is the diagonal matrix consisting of the feedback gains in the cluster $\Cluster_k$.
Similarly to \eqref{eq:original_phase_difference_dynamics}, by multiplying \eqref{eq:feedback_phase_dynamics} by $\tB_\intra$ from the left, 
we can derive the dynamics of $x^{(k)}(t)$ in \eqref{eq:def_x} with mean-phase feedback control as follows:
\begin{align}
\nonumber
    \begin{bmatrix}
        \dot{x}^{(1)}(t) \\ \vdots \\ \dot{x}^{(m)}(t)
    \end{bmatrix}
    =  
    &\begin{bmatrix}
        F_{\intra,x}^{(1)}(x^{(1)}(t)) \\ \vdots \\ F_{\intra,x}^{(m)}(x^{(m)}(t))
    \end{bmatrix}
    + F_{\inter,x}(x(t), z(t))\\ 
\label{eq:feedback_phase_difference_dynamics}
    &+
    \begin{bmatrix}
        F_{\FB,x}^{(1)}(x^{(1)}(t))\\
        \vdots\\
        F_{\FB,x}^{(m)}(x^{(m)}(t))
    \end{bmatrix},
\end{align}
where 
\begin{equation*}
    F_{\FB,x}^{(k)}(x^{(k)}(t)) := \tB_k^\T G_k\sin\left(-\frac{1}{|\Cluster_k|}L_{\FB,k}(\tB_k^\T)^+x^{(k)}(t)\right).
\end{equation*}
We have also used $\tB_\intra^\T\omega = 0$ under Assumption~\ref{asm:cluster_sync_asm}~(i).

\subsection{Design of Stabilizing Gains}
Here, let $g^{(k)}$ be the vector whose entries are the feedback gains to oscillators in the cluster $\Cluster_k$ (i.e., $\diag(g^{(k)}) = G_k$), and let $J_{\FB, k}(g^{(k)})$ be the Jacobian of $F_{\FB,x}^{(k)}(x^{(k)})$ at $x^{(k)} = 0$.
First, we compute $J_{\intra, k}$ and $J_{\FB, k}(g^{(k)})$ as follows:
\begin{align}
    \nonumber
    J_{\intra,k} &= \left.\frac{\partial F_{\intra,x}^{(k)}(x^{(k)})}{\partial x^{(k)}}\right|_{x^{(k)}=0} \\
    &= -\tB_k^\T B_kW_kB_k^\T(\tB_k^\T)^+,\\
    \nonumber
    J_{\FB,k}(g^{(k)}) &= \left.\frac{\partial F_{\FB,x}^{(k)}(x^{(k)})}{\partial x^{(k)}}\right|_{x^{(k)}=0} \\
    \label{eq:jacobian_feedback}
    &= -\frac{1}{|\Cluster_k|}\tB_k^\T \diag(g^{(k)})L_{\FB,k}(\tB_k^\T)^+,
\end{align}
where $(\tB_k^\T)^+$ is the Moore--Penrose inverse of $\tB_k^\T$.
We have also used $Q_k = B_k^\T(\tB_k^\T)^+$.

Let $L_{\intra,k}$ be the weighted Laplacian matrix of the graph $\Graph_k$, that is, $L_{\intra,k} := B_kW_kB_k^\T$.
Hence, letting $J_k(g^{(k)}) := J_{\intra,k} + J_{\FB,k}(g^{(k)})$, we obtain
\begin{equation}
\label{eq:feedback_jacobi_matrix}
    J_k(g^{(k)}) = - \tB_k^\T\left(L_{\intra,k} + \frac{1}{|\Cluster_k|}\diag(g^{(k)})L_{\FB,k}\right)(\tB_k^\T)^+.
\end{equation}
Note that the matrix in the big parentheses is the sum of the Laplacian matrix of $\Graph_k$ and that of the weighed complete graph, which depends on the feedback gains.
Therefore, in the vicinity of the equilibrium point, the inputs by mean-phase feedback strengthen the intra-cluster couplings.

The next theorem is our first main result of this paper.
It provides a sufficient condition to find the feedback gains for each cluster.
Note that in this result, the gains are constrained to be uniform within each cluster.
Later, we will present a numerical approach for finding the gains without imposing such a constraint.
\begin{thm}
\label{thm:feedback_feasibility}
    Consider the Kuramoto oscillators with mean-phase feedback control in \eqref{eq:feedback_Kuramoto_Oscillators}.
    Suppose that Assumption~\ref{asm:cluster_sync_asm} holds.
    For each $k \in \Set{\Range{1}{m}}$, take a scalar $\hat{g}_k$ sufficiently large that 
    \begin{equation}
        \label{eq:feedback_main_result}
        \hat{g}_k > \gamma^{(kk)} + \frac{1}{2}\lambda_{\rm max}(J_{\intra,k} + J^\T_{\intra, k}),
    \end{equation}
    and let $g^{(k)} = \hat{g}_k\ones_{|\Cluster_k|}$.
    Then, the cluster synchronization manifold is locally exponentially stable.
\end{thm}

\begin{proof}
    Here, we employ Proposition~\ref{prop:clusterwise_stability} and must show
\begin{equation}
    \label{eq:stability_condition_feedback}
    \lambda_{\max}(J_k(g^{(k)}) + J_k^\T(g^{(k)})) < -2\gamma^{(kk)}.
\end{equation}
    This condition is obtained by noticing that $J_k(g^{(k)})$ is the Jacobian for the term in \eqref{eq:feedback_phase_difference_dynamics} corresponding to the intra-cluster dynamics (as $J_{\intra, k}$ is for the similar terms in \eqref{eq:original_phase_difference_dynamics}).

    First, since $g^{(k)} = \hat{g}_k\ones_{|\Cluster_k|}$, the Jacobian $J_{\FB,k}(g^{(k)})$ in \eqref{eq:jacobian_feedback} can be written as
    \begin{equation*}
        J_{\FB,k}(g^{(k)}) = -\frac{\hat{g}_k}{|\Cluster_k|}\tB_k^\T L_{\FB,k}(\tB_k^\T)^+.
    \end{equation*}
    Note that $\tB_k^\T L_{\FB,k} = |\Cluster_k|\tB_k^\T$ as $L_{\FB,k}$ is the Laplacian matrix of the (unweighted) complete graph of order $|\Cluster_k|$.
    By noting that $\tB_k^\T$ has full row rank, we obtain
    \[
        J_{\FB,k}(g^{(k)}) = \hat{g}_k\tB_k^\T(\tB_k^\T)^+ = -\hat{g}_kI_{|\Cluster_k|-1}.
    \]
    Hence, $J_{\FB,k}(g^{(k)})$ is a diagonal matrix and $\eigmax(J_{\FB,k}(g^{(k)}) + J_{\FB,k}^\T(g^{(k)})) = -2\hat{g}_k$ is satisfied.
    From Weyl's inequality \cite{horn-matrix},
    \begin{align*}
        &\eigmax(J_k(g^{(k)})+J_k^\T(g^{(k)})) \\
        &\leq \eigmax(J_{\intra,k}+J_{\intra,k}^\T) + \eigmax(J_{\FB,k}(g^{(k)}) + J_{\FB,k}^\T(g^{(k)}))\\
        &= \eigmax(J_{\intra,k}+J_{\intra,k}^\T) - 2\hat{g}_k.
    \end{align*}
    By \eqref{eq:feedback_main_result}, we obtain \eqref{eq:stability_condition_feedback}.
\end{proof}

Theorem~\ref{thm:feedback_feasibility} indicates that stability can be guaranteed by using identical feedback gains which are large enough for all oscillators within each cluster.
However, the analytical result can be conservative, and it is clearly desirable to reduce the number of nodes that require feedback.
We next look at designing non-identical feedback gains by formulating an optimization problem.

\subsection{Design Method of Feedback Gains}
\label{subsec:optimization}
Here, our aim is to minimize the number of nodes to be controlled while achieving the stability of cluster synchronization.
Specifically, we formulate the feedback gain design as an optimization problem minimizing the $l_1$-norm of the gain.

To specify the nodes where control can be applied, we introduce the vector $h_k\in\Real^{|\Cluster_k|}$ defined by
\begin{equation*}
    (h_k)_i := \left\{
    \begin{array}{ll}
        0 & \text{if node $i$ can be controlled},\\
        1 & \text{if node $i$ cannot be controlled}.
    \end{array}
    \right.
\end{equation*}
To design the feedback gains, we consider the following optimization problem:
\begin{align}
\nonumber
    \underset{g^{(k)}}{\min}&\; \|g^{(k)}\|_1\\
\label{eq:feedback_constraint_stability}
    \subjectto&\; \eigmax(J_k(g^{(k)})+J_k^\T(g^{(k)})) + 2\gamma^{(kk)} + \epsilon \leq 0,\\
\label{eq:feedback_constraint_non-negative}
    &\; g^{(k)} \odot h_k = 0, \;\;\;  g^{(k)} \geq 0,
\end{align}
where $\odot$ indicates entrywise product of two vectors.
We emphasize that this is a convex problem and can be solved efficiently.
This can be confirmed as follows.
First, we used the $l_1$-norm, which represents the sum of the absolute values of entries of a vector and is moreover convex.
For obtaining a sparse solution, it is common to use $l_1$-norm instead of the $l_0$-norm, which in fact counts the number of nonzero entries of a vector, but is not a convex function.
Second, the transformation from the vector $g^{(k)}$ to the matrix $J_k(g^{(k)})+J_k^\T(g^{(k)})$ is affine (see \eqref{eq:feedback_jacobi_matrix}), and the function to obtain the maximum eigenvalue of a symmetric matrix is convex \cite{boyd-2004}.
Hence, the set of $g^{(k)}$ satisfying \eqref{eq:feedback_constraint_stability} and \eqref{eq:feedback_constraint_non-negative} is convex.
From Theorem~\ref{thm:feedback_feasibility}, it is clear that the feasible region is nonempty, if the vector $h_k$ has sufficiently many zero entries.

By solving the convex optimization problem above, we can find a possibly sparse set of nodes that are required to receive the feedback signals.
We will study our method numerically in Section~\ref{sec:numerical example}.

\section{Stabilization by Pacemakers}
\label{sec:pacemaker}
In this section, we consider the second approach for stabilization of the cluster synchronization manifold.
This one is based on inserting pacemakers in the network.
A pacemaker is an external input node that continuously oscillates at a constant frequency.
This approach was studied in \cite{doyle-2013} for the full synchronization case and is extended here for the cluster synchronization case.
As some clusters may not require such an additional input, 
we also investigate which clusters should be connected to pacemakers.

\subsection{Pacemakers}
Following \cite{doyle-2013}, we introduce an additional node, called the pacemaker, to each cluster $\Cluster_k$.
Its function is to keep the cluster synchronization manifold invariant by making all nodes in the cluster to follow its phase as follows.
Its phase is denoted by $\phi_k(t)$ and has a constant angular velocity equal to the natural frequency $\Omega_k$ of the nodes in the cluster as 
\[
\dot{\phi}_k(t) = \Omega_k.
\]
In cluster $\Cluster_k$, the pacemaker is connected to the nodes with a weight $v_k \geq 0$, and thus the phase of node $i$ is governed by
\begin{align}
    \nonumber
    \dot{\theta}_i(t) &= \Omega_k + \sum_{j=1}^n a_{ij}\sin(\theta_j(t) - \theta_i(t))\\
    \label{eq:kuramoto_oscillator_with_pacemakers}
    &\hspace{1em} + v_k\sin(\phi_k(t) - \theta_i(t)),\:\:\:\:i\in\Cluster_k.
\end{align}
We point out that the pacemaker maintains its own angular velocity at $\Omega_k$, but the nodes connected to it may not oscillate at this frequency unless the weight $v_k$ is very large.
This aspect introduces some differences in the analysis from the previous case with mean-phase feedback control as we will see.

Defining the phase difference between nodes $i$ and $j$ in $\Cluster_k$ by $x_{ij}(t) := \theta_j(t) - \theta_i(t)$, we obtain
\begin{align}
    \nonumber
    \dot{x}_{ij}(t) =& \sum_{h=1}^n \left[ a_{jh}\sin(x_{jh}(t)) - a_{ih}\sin(x_{ih}(t)) \right] \\
    \label{eq:pacemaker_phase_difference}
    & + v_k\left[\sin(\phi_k(t) - \theta_j(t)) - \sin(\phi_k(t) - \theta_i(t))\right].
\end{align}
In particular, the second term on the right-hand side is the input from the pacemaker. 
By a trigonometric identity, this term can be expressed as
\begin{align}
\nonumber
    &v_k\left[\sin(\phi_k(t) - \theta_j(t)) - \sin(\phi_k(t) - \theta_i(t))\right]\\
\label{eq:pacemaker_element}
    &= -2v_k\cos\left(\frac{x_{ij}(t)-2\xi_{kj}(t)}{2}\right)\sin\left(\frac{x_{ij}(t)}{2}\right),
\end{align}
where $\xi_{kj}(t) := \phi_k(t) - \theta_j(t)$ is the phase difference between the pacemaker of the cluster $k$ and the oscillator $i$. To simplify the notation, let $\xi^{(k)}(t) := [\xi_{ki}(t)]_{i\in\Cluster_k}$.
Then, we define the vector $F_{\PM,x}^{(k)}(x^{(k)}(t),\xi^{(k)}(t))$ whose entries correspond to \eqref{eq:pacemaker_element} for nodes in $\Cluster_k$.
Note that it is a function of only $x^{(k)}$ and $\xi^{(k)}$.
Finally, using \eqref{eq:original_phase_difference_dynamics}, we obtain the phase difference dynamics in the presence of pacemakers as
\begin{align}
\nonumber
    \begin{bmatrix}
        \dot{x}^{(1)}(t) \\ \vdots \\ \dot{x}^{(m)}(t)
    \end{bmatrix} 
    =\; & 
    \begin{bmatrix}
        F_{\intra,x}^{(1)}(x^{(1)}(t)) \\ \vdots \\ F_{\intra,x}^{(m)}(x^{(m)}(t))
    \end{bmatrix}
    + F_{\inter,x}(x(t), z(t)) \\
\label{eq:pacemaker_phase_difference_dynamics}
    &+ 
    \begin{bmatrix}
        F_{\PM,x}^{(1)}(x^{(1)}(t),\xi^{(1)}(t)) \\ \vdots \\ F_{\PM,x}^{(m)}(x^{(m)}(t),\xi^{(m)}(t))
    \end{bmatrix}.
\end{align}

\subsection{Design of Stabilizing Weights}
The following lemma provides a condition for the phase differences among nodes in the same cluster to remain within given bounds determined by the pacemaker weight.
To this end, let 
\begin{equation}
    \label{eq:pacemaker_inv_set}
    \Xi_{\rm inv}^{(k)} := \Set{\Comp{\xi^{(k)}}{\xi_{ki} \in[-\epsilon_k(v_k),\epsilon_k(v_k)],\,i\in\Cluster_k}},
\end{equation}
where for $v_k$ satisfying $v_k \geq \sum_{l=1,l\neq k}^m\sum_{j\in\Cluster_l} a_{ij}$, we set
\begin{equation}
    \label{eq:epsilon}
    \epsilon_k(v_k) := \sin^{-1}\left(\frac{\sum_{l=1,l\neq k}^m\sum_{j\in\Cluster_l} a_{ij}}{v_k}\right).
\end{equation}
\begin{lem}
\label{lem:pacemaker_invariance}
    For $k\in\Set{\Range{1}{m}}$, if $v_k \geq \sum_{l=1,l\neq k}^m\sum_{j\in\Cluster_l} a_{ij}$,
    then the set $\Xi_{\rm inv}^{(k)}$ in \eqref{eq:pacemaker_inv_set} is forward invariant.
\end{lem}
\begin{proof}
    It can be observed from \eqref{eq:pacemaker_element} that if 
    \(
        -\frac{\pi}{2} < \phi_k(t) - \frac{\theta_j(t) + \theta_i(t)}{2} < \frac{\pi}{2},
    \)
    then $x_{ij}(t)$ has the same sign as 
    \(
        \cos\left(\frac{x_{ij}(t)-2\xi_{kj}(t)}{2}\right)\sin\left(\frac{x_{ij}(t)}{2}\right).
    \)
    The latter means that the pacemaker reduces the phase difference between the oscillators $i$ and $j$.
    Let $i = \argmax_{i'\in\Cluster_k} |\xi_{ki'}|$. 
    Then, we must show that $\xi_{ki} \geq \epsilon_k(v_k)$ implies $\dot{\xi}_{ki} \leq 0$,
    and $\xi_{ki} \leq -\epsilon_k(v_k)$ implies $\dot{\xi}_{ki} \geq 0$.
    When $\xi_{ki} \geq \epsilon_k(v_k)$, the time derivative of $\xi_{ki}$ can be upper bounded as
    \begin{align*}
        \dot{\xi}_{ki} &= \omega_i - \Omega_k + \sum_{j\in\Cluster_k}a_{ij}\sin(\theta_j - \theta_i) \\
        &\mbox{\hspace{1.5em}} + \sum_{l=1,\,l\neq k}^m\sum_{j\in\Cluster_l}a_{ij}\sin(\theta_j - \theta_i) + v_k\sin(\phi_k - \theta_i)\\
        &= \sum_{j\in\Cluster_k}a_{ij}\sin(\xi_{kj} - \xi_{ki}) \\
        &\mbox{\hspace{1.5em}}+ \sum_{l=1,\,l\neq k}^m\sum_{j\in\Cluster_l}a_{ij}\sin(\theta_j - \theta_i) - v_k\sin(\xi_{ki})\\
        &\leq \sum_{l=1,\,l\neq k}^m\sum_{j\in\Cluster_l}a_{ij} - v_k\sin(\xi_{ki}).
    \end{align*}
    Therefore, by \eqref{eq:epsilon}, $\xi_{ki} \geq \epsilon_k(v_k)$ implies $\dot{\xi}_{ki} \leq 0$.
    We can similarly show that if $\xi_{ki} \leq -\epsilon_k(v_k)$, then $\dot{\xi}_{ki} \geq 0$.
\end{proof}
This lemma indicates that by increasing the pacemaker weight, the maximum phase difference between the pacemaker and oscillators in a cluster can be made arbitrarily small.

We are ready to state our main result for cluster synchronization based on pacemakers.
To this end, let
\begin{equation}
\label{eq:pacemaker_constant}
    y_k := \frac{1}{2}\eigmax(J_{\intra,k} + J_{\intra,k}^\T) + \gamma^{(kk)}
\end{equation}
for $k\in\Set{\Range{1}{m}}$.
The following theorem demonstrates that cluster synchronization occurs if the pacemaker weights are chosen to be sufficiently large.
Furthermore, pacemakers are needed in clusters whose corresponding $y_k$ are positive.
\begin{thm}
\label{thm:pacemaker_gain}
    Consider the Kuramoto oscillators with pacemakers in \eqref{eq:kuramoto_oscillator_with_pacemakers}. 
    Suppose that Assumption~\ref{asm:cluster_sync_asm} holds.
    For every $k$ such that $y_k > 0$, take $v_k$ large enough that 
    $y_k < v_k\cos(\epsilon_k(v_k))$ and $v_k \geq \sum_{l=1,l\neq k}^m\sum_{j\in\Cluster_l} a_{ij}$.
    Then, the cluster synchronization manifold is locally exponentially stable.
    Specifically, if $\xi^{(k)}(0) \in \Xi_{\rm inv}^{(k)}$, then $\theta(t)$ reaches cluster synchronization.
\end{thm}
\begin{proof}
    Suppose $\xi^{(k)}(0) \in \Xi_{\rm inv}^{(k)}$ for all $k\in\Set{\Range{1}{m}}$.
    From Lemma~\ref{lem:pacemaker_invariance}, the Jacobian matrix of $F_{\PM,x}^{(k)}(x^{(k)},\xi^{(k)})$ in \eqref{eq:pacemaker_phase_difference_dynamics} at the equilibrium $x = 0$ is
    \begin{align*}
        J_{\PM,k}(v_k) &:= \left.\frac{\partial F_{\PM,x}^{(k)}(x^{(k)},\xi^{(k)})}{\partial x^{(k)}}\right|_{x^{(k)}=0}\\ 
        &= -v_k\diag\left(\{\cos(\xi_{ki})\}_{(i,j)\in\tilde{\Edges}_k}\right)\\
        &= -v_k(\cos(\epsilon_k(v_k))I + D_k),
    \end{align*}
    where 
    $D_k := \diag(\Set{\cos(\xi_{ki})-\cos(\epsilon_k(v_k))}_{(i,j)\in\tilde{\Edges}_k})$.
    Note that by assumption and Lemma~\ref{lem:pacemaker_invariance}, $D_k \geq 0$.
    We let $J_k(v_k) := J_{\intra, k} + J_{\PM,k}(v_k)$. 
    Under Assumption~\ref{asm:cluster_sync_asm}, we use the solution $X_k$ of \eqref{eq:lyapunov_eq}.
    Then, for any $x_k\neq 0$, it holds
    \begin{align*}
        x_k^\T(D_k^\T X_k + X_kD_k)x_k &= 2(D_kx_k)^\T X_kx_k\\
        &\leq -2 \min_i [D_k]_{ii} x_k^\T X_kx_k \leq 0,
    \end{align*}
    where the minimum is taken over the diagonal entries of $D_k$.
    Now, it is straightforward to show that with $X_k$ from \eqref{eq:lyapunov_eq}, it holds
    $J_k(v_k)^\T X_k + X_kJ_k(v_k) \preceq -I$.
    This means that we can apply Proposition~\ref{prop:clusterwise_stability} to the system in \eqref{eq:kuramoto_oscillator_with_pacemakers}.
    Thus, we must show
    \begin{equation}
        \label{eq:pacemaker_cs_condition}
        \lambda_{\max}(J_k(v_k) + J_k^\T(v_k)) < -2\gamma^{(kk)}.
    \end{equation}
    From Weyl's inequality \cite{horn-matrix}, we obtain
    \begin{align*}
         &\eigmax(J_k(v_k) + J_k^\T(v_k)) \\
        &= \eigmax(J_{\intra,k} + J_{\intra,k}^\T + J_{\PM,k}(v_k) + J_{\PM,k}^\T(v_k))\\
        &= \eigmax(J_{\intra,k} + J_{\intra,k}^\T - 2v_k(\cos(\epsilon_k(v_k))I + D_k))\\
        &\leq \eigmax(J_{\intra,k} + J_{\intra,k}^\T - 2v_k\cos(\epsilon_k(v_k))I)\\
        &= \eigmax(J_{\intra,k} + J_{\intra,k}^\T) -2v_k\cos(\epsilon_k(v_k)).
    \end{align*}
    Therefore, by $y_k < v_k\cos(\epsilon_k(v_k))$, we arrive at \eqref{eq:pacemaker_cs_condition}.
\end{proof}
The pacemakers may be relatively easy to design and implement as this approach is based on feedforward control.
The design requires the knowledge of the natural frequency of the cluster to which it will be applied.
In practice, this may be slightly difficult, especially in networks where the natural frequencies can be different among the nodes in one cluster and may even fluctuate over time.
We will consider such cases in the next section.

Note that we are only interested in synchronization between nodes in each cluster, so the nodes and the pacemaker are not necessarily synchronized.  Observe that in \eqref{eq:kuramoto_oscillator_with_pacemakers}, even under cluster synchronization, the term $v_k\sin(\phi_k - \theta_i)$ may be nonzero, but under uniform weights, in \eqref{eq:pacemaker_phase_difference}, the second term on the right-hand side is zero. This means that the pacemaker weights for each cluster must be uniform at each node for invariance of the cluster synchronization manifold. 

\section{Cluster Phase Cohesiveness}
\label{sec:cluster_phase_cohesiveness}
So far, we have focused on attaining cluster synchronization for Kuramoto oscillators in a strict sense so that in each cluster, all phases asymptotically synchronize without any error.
In this section, we consider relaxing this goal where it is enough for the clusters to reach approximate synchronization.
This is called cluster phase cohesiveness, and we study how to achieve this via mean-phase feedback control and pacemakers.

\subsection{Problem Setup}
In this section, we deal with Kuramoto oscillators for which cluster synchronization states are not invariant.
Hence, we consider the case where the conditions in Assumption~\ref{asm:cluster_sync_asm} do not hold.
To this end, for $k,l\in\Set{\Range{1}{m}}$ and $k\neq l$, let
\begin{align}
    \label{eq:diff_freq}
    \Delta\omega_k &:= \max_{i,j\in\Cluster_k}|\omega_i-\omega_j|,\\
    \label{eq:diff_eep}
    \epsilon_{kl} &:= \max_{i\in\Cluster_k}\sum_{j\in\Cluster_l}a_{ij} - \min_{i\in\Cluster_k}\sum_{j\in\Cluster_l}a_{ij}.
\end{align}
It is clear that $\Delta\omega_k = 0$ for all $k$ if and only if Assumption~\ref{asm:cluster_sync_asm}~(i) holds. Moreover, $\epsilon_{kl} = 0$ for all $k,l$ if and only if Assumption~\ref{asm:cluster_sync_asm}~(ii) holds.

We however would like the oscillators to achieve a certain level of cluster synchronization with a bounded error for all clusters.
We specify the desired level of error by $\psi\in[0,\pi]$.
Then, we define the cluster phase cohesiveness set $\CSM(\psi)$ with respect to the partition $\Pi$ by
\[
    \CSM(\psi) := \left\{\theta \in \mathbb{T}^n \;\left|\; \max_{i,j\in\Cluster_k} |\theta_i - \theta_j| \leq \psi,\,k\in\{1,\,\dots\,,m\}\right\}\right..
\]

For the stability analysis of cluster phase cohesiveness, we follow a Lyapunov approach \cite{cao-2021, lin-2007}. 
Specifically, let 
\[
    V_k(\theta^{(k)}) := \max_{i,j\in\Cluster_k} |\theta_i - \theta_j| = \|B_{{\rm comp},k}^\T\theta^{(k)}\|_\infty
\]
for $k\in\{1,\,\dots\,,m\}$, where $B_{{\rm comp},k}$ is the incidence matrix of the complete graph within the $\Cluster_k$.
The upper Dini derivative of $V_k(\theta^{(k)}(t))$ is given by
\begin{align*}
    \mathrm{D}^+_tV_k(\theta^{(k)}(t)) &:= \limsup_{\tau\downarrow0}\frac{V_k(\theta^{(k)}(t + \tau)) - V_k(\theta^{(k)}(t))}{\tau}\\
    &= \dot{\theta}_{i_k}(t) - \dot{\theta}_{j_k}(t),\;\;\;\;\;\forall (i_k,j_k)\in\mathcal{I}_k(t),
\end{align*}
where
\begin{align*}
    \mathcal{I}_k(t) := \{ &(i,j) \in \mathcal{I}'_k(t) \mid \\
    &\dot{\theta}_i(t) - \dot{\theta}_j(t) = \max_{i',j'\in\Cluster_k} (\dot{\theta}_{i'}(t) - \dot{\theta}_{j'}(t))\},\\
    \mathcal{I}'_k(t) := \{ &(i,j)\in\Cluster_k\times\Cluster_k\mid \\
    &|\theta_i(t) - \theta_j(t)| = \max_{i',j'\in\Cluster_k}|\theta_{i'}(t) - \theta_{j'}(t)|\}.
\end{align*}
By the definition of $V_k$, we can write $\CSM(\psi) = \{\theta\in\mathbb{T}^n\mid V_k(\theta^{(k)})\leq \psi_,\,k\in\{1,\,\dots\,,m\}\}$.
It is now clear that the Kuramoto oscillators for which Assumption~\ref{asm:cluster_sync_asm} does not hold, cluster phase cohesiveness can be reached, that is, $\CSM(\psi)$ is a positively invariant set if $\mathrm{D}_t^+V_k(\theta^{(k)})\leq 0$ for any $\theta^{(k)}$ such that $V_k(\theta^{(k)}) = \psi$.

\subsection{Stabilization by Mean-phase Feedback Control}
We define the three functions $f_{\intra,k}:[0,\pi]\rightarrow\Real$, $f_{\inter,k}:[0,\pi]\rightarrow\Real$, and $f_{\FB,k}:[0,\pi]\rightarrow\Real$ as follows:
\begin{align*}
    f_{\intra,k}(\psi) &:= \underline{a}_k\underline{d}_k\sin\psi,\\
    f_{\inter,k}(\psi) &:= \min(2D_{\inter,k}, 2D_{\inter,k}\psi + \epsilon_k),\\
    f_{\FB, k}(\psi) &:= -\bar{g}_k\sin\psi + 2(\bar{g}_k - \underline{g}_k),
\end{align*}
where 
\begin{align*}
    \underline{a}_k &:= \min_{i,j\in\Cluster_k,a_{ij} \neq 0} a_{ij},\,\,\,\, \underline{d}_k := \min_{i\in\Cluster_k} |\mathcal{D}_{k,i}|,\\
    \mathcal{D}_{k,i} &:= \Set{\{i',j'\}\,\mid\,i',j'\in\Cluster_k,\, i'\neq j',\,\\
    &\hspace{6em} a_{ii'} > 0,\,a_{ij'} > 0,\,a_{i'j'}>0},\\
    D_{\inter,k} &:= \max_{i\in\Cluster_k}\sum_{l=1,l\neq k}^m\sum_{j\in\Cluster_l}a_{ij},\\
    \bar{g}_k &:= \max_{i\in\Cluster_k} g_i,\,\,\,\, \underline{g}_k := \min_{i\in\Cluster_k} g_i.
\end{align*}
The set $\mathcal{D}_{k,i}$ denotes the set of unordered pairs of adjacent nodes in cluster $k$ which have node $i$ as a common neighbor.

The next theorem provides a sufficient condition for cluster phase cohesiveness via mean-phase feedback control.
\begin{thm}
    \label{thm:CPC_via_mean-phase_feedback_control}
    Consider the Kuramoto oscillators with mean-phase feedback control in \eqref{eq:feedback_Kuramoto_Oscillators}.
    Let $\underline{\psi} \in (0, \frac{\pi}{2})$. For each $k\in\Set{\Range{1}{m}}$, take $\hat{g}_k$ sufficiently large that 
    \begin{equation}
        \label{eq:cpc_feedback_main_result}
            \hat{g}_k \geq \frac{\Delta\omega_k - f_{\intra,k}(\underline{\psi}) + f_{\inter,k}(\underline{\psi})}{\sin\underline{\psi}},
    \end{equation}
    and let $g^{(k)} = \hat{g}_k\ones_{|\Cluster_k|}$. Then, there exists a constant $\bar{\psi} \in (\frac{\pi}{2},\pi)$ such that the following properties hold:
    \begin{enumerate}
        \item The set $\CSM(\psi)$ is positively invariant for any $\psi \in [\underline{\psi},\bar{\psi}]$.
        \item Any solution $\theta^{(k)}(t)$ whose initial state satisfies $\underline{\psi} < V_k(\theta^{(k)}(0)) < \bar{\psi}$ reaches $\CSM(\underline{\psi})$.
    \end{enumerate}
\end{thm}
\begin{proof}
    From the definition \eqref{eq:diff_freq} of $\Delta\omega_k$, we have $\omega_{i_k} - \omega_{j_k} \leq \Delta\omega_k$ for any $(i_k, j_k) \in \mathcal{I}_k(t)$.
    We fix a time $t$ and let $\theta_{i_k} - \theta_{j_k} = \psi$ for any $(i_k, j_k) \in \mathcal{I}_k(t)$ and for any $k\in\{1,\,\dots\,,m\}$.
    For the dynamics of Kuramoto oscillators \eqref{eq:feedback_Kuramoto_Oscillators}, 
    \begin{align}
    \nonumber
    &\mathrm{D}^+_tV_k(\theta^{(k)}(t)) = \dot{\theta}_{i_k} - \dot{\theta}_{j_k}\\
    \nonumber
    &= \omega_{i_k} - \omega_{j_k} + \sum_{h=1}^n a_{i_kh}\sin(\theta_h - \theta_{i_k}) - \sum_{h=1}^n a_{j_kh}\sin(\theta_h - \theta_{j_k})\\
    \nonumber
    &= \omega_{i_k} - \omega_{j_k} + \sum_{h\in\Cluster_k}[a_{i_kh}\sin(\theta_h - \theta_{i_k}) - a_{j_kh}\sin(\theta_h - \theta_{j_k})]\\
    \nonumber
    &\mbox{\hspace{1.5em}} + \sum_{l = 1, l \neq k}^m \sum_{h\in\Cluster_l}[a_{i_kh}\sin(\theta_h - \theta_{i_k}) - a_{j_kh}\sin(\theta_h - \theta_{j_k})]\\
    \label{eq:CPC_diff_phase}
    &\mbox{\hspace{1.5em}} + g_{i_k}\sin(\theta_{\av,k} - \theta_{i_k}) - g_{j_k}\sin(\theta_{\av, k} - \theta_{j_k})
    \end{align}
    In \eqref{eq:CPC_diff_phase}, on the far right-hand side, the third term corresponds to the intra-cluster interactions, the fourth term is the inter-cluster interactions, and the fifth term is the mean-phase feedback control.
    We first start by evaluating the terms in \eqref{eq:CPC_diff_phase} corresponding to the intra-cluster interactions as 
    \begin{align}
        \nonumber
        &\sum_{h\in\Cluster_k}[a_{i_kh}\sin(\theta_h - \theta_{i_k}) - a_{j_kh}\sin(\theta_h - \theta_{j_k})] \\
        \nonumber
        &\leq \sum_{h\in\Cluster_k,\;a_{i_kh},a_{j_kh} \neq 0}[a_{i_kh}\sin(\theta_h - \theta_{i_k}) - a_{j_kh}\sin(\theta_h - \theta_{j_k})]\\
        \label{eq:CPC_intra_cluster_1_1}
        &\leq -\underline{a}_k\sum_{h\in\Cluster_k,\;a_{i_kh},a_{j_kh} \neq 0} [\sin(\theta_{i_k} - \theta_h) - \sin(\theta_{j_k} - \theta_h)].
    \end{align}
    By applying a trigonometric identity to the summand term on the far right-hand side, we obtain
    \begin{align}
        \nonumber
        &\sin(\theta_{i_k} - \theta_h) - \sin(\theta_{j_k} - \theta_h) \\
        \label{eq:CPC_tri}
        &= 2\sin\left(\frac{\psi}{2}\right)\cos\left(\frac{\theta_{i_k} - \theta_h}{2} - \frac{\theta_h - \theta_{j_k}}{2}\right).
    \end{align}
    From the choices of $i_k$ and $j_k$, we have
    \[
        -\frac{\psi}{2} \leq \frac{\theta_{i_k} - \theta_h}{2} - \frac{\theta_h - \theta_{j_k}}{2} \leq \frac{\psi}{2},
    \]
    for any $h \in \Cluster_k$. From these inequalities,
    \[
        \cos\left(\frac{\theta_{i_k} - \theta_h}{2} - \frac{\theta_h - \theta_{j_k}}{2}\right) \geq \cos\left(\frac{\psi}{2}\right).
    \]
    Thus, we have from \eqref{eq:CPC_tri}
    \begin{align*}
        \sin(\theta_{i_k} - \theta_h) - \sin(\theta_{j_k} - \theta_h) &\geq 2\sin\left(\frac{\psi}{2}\right)\cos\left(\frac{\psi}{2}\right) \\
        &= \sin(\psi).
    \end{align*}
    By using this relation and \eqref{eq:CPC_intra_cluster_1_1}, we obtain
    \[
        \sum_{h\in\Cluster_k}[a_{i_kh}\sin(\theta_h - \theta_{i_k}) - a_{j_kh}\sin(\theta_h - \theta_{j_k})] \leq -\underline{a}_k\underline{d}_k\sin(\psi).
    \]

    Next, we move on to the term for inter-cluster interactions in \eqref{eq:CPC_diff_phase}, which can be evaluated as
    \begin{align}
        \nonumber
        &\sum_{l = 1, l \neq k}^m \sum_{h \in \Cluster_l} \left[a_{{i_k} h} \sin(\theta_h - \theta_{i_k}) - a_{{j_k} h} \sin(\theta_h - \theta_{j_k})\right] \\
        \label{eq:CPC_inter_cluster_1}
        &\leq \sum_{l = 1, l \neq k}^m \sum_{h \in \Cluster_l}[a_{i_kh} + a_{j_kh}] \leq 2D_{\inter,k}.
    \end{align} 
    On the other hand, using $\theta_{i_l}$ such that $\theta_{i_l} - \theta_{j_l} = \psi$ for some $j_l$, we have
    \begin{align*}
        \sin(\theta_h - \theta_{i_k}) &= \sin(\theta_h - \theta_{i_l} + \theta_{i_l} - \theta_{i_k} + \theta_{i_k} - \theta_{i_k}),\\
        \sin(\theta_h - \theta_{j_k}) &= \sin(\theta_h - \theta_{i_l} + \theta_{i_l} - \theta_{i_k} + \theta_{i_k} - \theta_{j_k}).
    \end{align*} 
    Moreover, note that for any $a, b \in \mathbb{R}$, there exists a constant $\delta$ such that $\sin(a + b) = \sin(a) + \delta$ for any $a, b \in \mathbb{R}$ with $|\delta| \leq |b|$. Hence, we have
    \begin{align*}
        \sin(\theta_h - \theta_{i_k}) &= \sin(\theta_{i_l} - \theta_{i_k}) + \delta_{i_kh},\\
        \sin(\theta_h - \theta_{j_k}) &= \sin(\theta_{i_l} - \theta_{i_k}) + \delta_{j_kh},
    \end{align*}
    where $|\delta_{i_kh}|\leq |\theta_h - \theta_{i_l}| \leq \psi$, $|\delta_{j_kh}| \leq |\theta_h - \theta_{i_l} + \psi| \leq \psi$.
    Therefore, using these relations and \eqref{eq:CPC_inter_cluster_1}, we obtain
    \begin{align*}
        &\sum_{l = 1, l \neq k}^m \sum_{h \in \Cluster_l} \left[a_{{i_k} h} \sin(\theta_h - \theta_{i_k}) - a_{{j_k} h} \sin(\theta_h - \theta_{j_k})\right]\\
        &= \sum_{l = 1, l \neq k}^m \sum_{h \in \Cluster_l} \left[a_{{i_k} h}(\sin(\theta_{i_l} - \theta_{i_k}) + \delta_{i_kh})\right. \\
        &\left.\hspace{22mm}- a_{j_k h}(\sin(\theta_{i_l} - \theta_{i_k}) + \delta_{j_kh})\right]\\
        &= \sum_{l = 1, l \neq k}^m \sum_{h \in \Cluster_l}(a_{i_kh} - a_{j_kh})\sin(\theta_{i_l} - \theta_{i_k}) \\
        &\mbox{\hspace{2em}} + \sum_{l = 1, l \neq k}^m \sum_{h \in \Cluster_l} (a_{i_kh}\delta_{i_kh} - a_{j_kh}\delta_{j_kh})\\
        &\leq \sum_{l = 1, l \neq k}^m \epsilon_{kl} + \sum_{l = 1, l \neq k}^m \sum_{h \in \Cluster_l}(a_{i_kh}|\delta_{i_kh}| + a_{j_kh}|\delta_{j_kh}|)\\
        &\leq \epsilon_k + 2D_{m, k}^\inter\psi.
    \end{align*}
    
    Finally, we evaluate the term of the mean-phase feedback control in \eqref{eq:CPC_diff_phase}. 
    This can be done as 
    \begin{align*}
        &g_{i_k}\sin(\theta_{\av,k} - \theta_{i_k}) - g_{j_k}\sin(\theta_{\av, k} - \theta_{j_k}) \\
        &= -\bar{g}_k[\sin(\theta_{i_k} - \theta_{\av, k}) - \sin(\theta_{j_k} - \theta_{\av, k})] \\
        &\mbox{\hspace{1em}} + (g_{i_k} - \bar{g}_k)\sin(\theta_{i_k} - \theta_{\av, k})\\
        &\mbox{\hspace{1em}} - (g_{j_k} - \bar{g}_kz)\sin(\theta_{j_k} - \theta_{\av, k})\\
        &\leq -\bar{g}_k\sin(\psi) + 2(\bar{g}_k - \underline{g}_k) = f_{\FB,k}(\psi).
    \end{align*}
    Therefore, the upper Dini derivative of the maximum phase difference is bounded from above as
    \[
        \dot{\theta}_{i_k} - \dot{\theta}_{j_k} \leq \Delta\omega_k - f_{\intra,k}(\psi) + f_{\inter,k}(\psi) + f_{\FB,k}(\psi).
    \]
    The function $-f_{\intra,k}(\psi) + f_{\FB, k}(\psi)$ is convex, and the function $\Delta\omega_k + f_{\inter,k}(\psi)$ is concave.
    Hence, when $g^{(k)} = \hat{g}_k\ones_{|\Cluster_k|}$, the equation $\Delta\omega_k - f_{\intra,k}(\psi) + f_{\inter,k}(\psi) + f_{\FB,k}(\psi) = 0$ 
    has two different solutions 
    $\underline{\psi}',\,\bar{\psi}'$ with $\underline{\psi}' \leq \underline{\psi}$, $\bar{\psi} \leq \bar{\psi}'$.
    In particular, in \eqref{eq:cpc_feedback_main_result}, if the equal sign holds, then $\underline{\psi}' = \underline{\psi}$ and $\bar{\psi}' = \bar{\psi}$ hold.
    Therefore, $\Delta\omega_k - f_{\intra,k}(\psi) + f_{\inter,k}(\psi) + f_{\FB,k}(\psi) \leq 0$ for any $\psi \in [\underline{\psi}, \bar{\psi}]$.
    Moreover, $\Delta\omega_k - f_{\intra,k}(\psi) + f_{\inter,k}(\psi) + f_{\FB,k}(\psi) < 0$ for any $\psi \in (\underline{\psi}, \bar{\psi})$. This completes the proof.
\end{proof}

Theorem~\ref{thm:CPC_via_mean-phase_feedback_control} provides an estimation of the region of attraction of the cluster phase cohesiveness set.
If larger errors are tolerated, the feedback gains can be reduced.
The difference between Theorem~\ref{thm:CPC_via_mean-phase_feedback_control} and the cluster phase cohesiveness condition in Theorem~2 of \cite{cao-2021} is the use of $\epsilon_{kl}$, which represents the error between the weights satisfying the EEP condition Assumption~\ref{asm:cluster_sync_asm}~(ii).
This makes the condition less conservative.
Theorem~\ref{thm:CPC_via_mean-phase_feedback_control} can also be applied when Assumption~\ref{asm:cluster_sync_asm} is satisfied, and cluster phase cohesiveness is possible with smaller gains when Assumption~\ref{asm:cluster_sync_asm} is not satisfied.
Since the condition of Theorem~\ref{thm:CPC_via_mean-phase_feedback_control} uses only the maximum and minimum of feedback gains for each cluster, optimization of feedback gains based on this theorem is difficult and is left for future research. 

\subsection{Stabilization by Pacemakers}
We define the function $f_{\PM,k}:[0, \pi]\rightarrow\Real$ by
\[
    f_{\PM,k}(\psi) := -\bar{v}_k\sin\psi + 2(\bar{v}_k - \underline{v}_k),
\]
where $\bar{v}_k$ and $\underline{v}_k$ are the maximum and the minimum pacemaker weights in $\Cluster_k$, respectively.
Let $\psi\in[0,\pi]$ be the constant of the error bound for all clusters. Then, we define the cluster phase cohesiveness set considering the phase difference from the pacemakers by
\begin{align*}
\CSM^\PM(\psi) := &\left\{ (\theta,\phi)\in \Torus^{n + m} \mid \max_{i,j\in\Cluster_k}|\theta_i - \theta_j| \leq \psi_k,\, \right.\\
&\hspace{3mm} \left.\max_{i\in\Cluster_k}|\theta_i - \phi_k| \leq \psi_k,\,k\in\{1,\,\dots\,,m\}\right\}.
\end{align*}
The next theorem provides a sufficient condition for cluster phase cohesiveness via pacemakers.
\begin{thm}
    \label{thm:CPC_via_pacemakers}
    Consider the Kuramoto oscillators with pacemakers in \eqref{eq:kuramoto_oscillator_with_pacemakers}.
    Let $\underline{\psi} \in (0, \frac{\pi}{2})$. For each $k\in\Set{\Range{1}{m}}$, take $\hat{v}_k$ sufficiently large that 
    \begin{equation}
        \label{eq:cpc_pacemaker_main_result}
            \hat{v}_k \geq \frac{\Delta\omega_k - f_{\intra,k}(\underline{\psi}) + f_{\inter,k}(\underline{\psi})}{\sin\underline{\psi}},
    \end{equation}
    and let $v_k = \hat{v}_k\ones_{|\Cluster_k|}$, $\Omega_k = \frac{1}{|\Cluster_k|}\sum_{i\in\Cluster_k}\omega_i $. Then, there exists a constant $\bar{\psi} \in (\frac{\pi}{2},\pi)$ such that the following properties hold:
    \begin{enumerate}
        \item The set $\CSM^\PM(\psi)$ is positively invariant for any $\psi \in [\underline{\psi},\bar{\psi}]$.
        \item Any solution $\theta^{(k)}(t)$ whose initial state satisfies $\underline{\psi} < V_k(\theta^{(k)}(0)) < \bar{\psi}$ and $\underline{\psi} < \max_{i\in\Cluster_k} \xi_{ki}(0) < \bar{\psi}$ reaches $\CSM^\PM(\underline{\psi})$.
    \end{enumerate}
\end{thm}
\begin{proof}
    We let $\xi_i(t) := \theta_i(t) - \phi_k(t)$. Then, the time derivative of $\xi_i(t)$ is
    \begin{align*}
        \dot{\xi}_i(t) &= \dot{\theta}_i(t) - \dot{\phi}_k(t) = \omega_i - \Omega_k + \sum_{j=1}^na_{ij}\sin(\theta_j(t) - \theta_i(t)) \\
        &\mbox{\hspace{11.1em} }+ v_i\sin(\phi_k(t) - \theta_i(t)).
    \end{align*}
    Here, we fix the time $t$ and the index $i_k \in \Cluster_k$ such that $\xi_{i_k}$ is the maximum phase difference at $t$.
    Also, let $\xi_i = \psi$. Then, 
    \begin{equation*}
        \dot{\xi}_i \leq \Delta\omega_k - f_{\intra,k}(\psi) + f_{\inter,k}(\psi) - f_{\PM,k}(\psi)
    \end{equation*}
    for $i\in\Cluster_k$.
    As in the proof of Theorem~\ref{thm:CPC_via_mean-phase_feedback_control}, the upper Dini derivative of the maximum phase difference can be  evaluated by 
    \[
        \dot{\theta}_{i_k} - \dot{\theta}_{j_k} \leq \Delta\omega_k - f_{\intra, k}(\psi) + f_{\inter, k}(\psi) + f_{\PM, k}(\psi).
    \]
    The proof is then completed by following an argument as that in Theorem~\ref{thm:CPC_via_mean-phase_feedback_control}.
\end{proof}

Theorem~\ref{thm:CPC_via_pacemakers} shows that the phase difference between the nodes and the pacemaker can be suppressed with the same bounds.
Similar to the mean-phase feedback control, if larger errors are tolerated, the pacemaker weights can be reduced.

\section{Discussion}
In this paper, we proposed two control methods for stabilizing cluster synchronization and cluster phase cohesiveness in Kuramoto oscillators: one based on periodic inputs generated by pacemakers, and the other based on feedback of the mean phase within each cluster. Here, we summarize and compare the characteristics of these methods in terms of control design, implementation, and performance.

The pacemaker generates a periodic input with a fixed frequency that matches the natural frequency of each cluster. This feedforward method does not require any measurement of the system state and thus can be implemented without sensing infrastructure. The control input is applied uniformly to all nodes within each cluster, which simplifies the input structure but may limit the flexibility in handling heterogeneous networks or node-specific constraints. Since the input signal is fixed a priori, the achievable synchronization pattern is also predetermined. Furthermore, the lack of feedback makes the method potentially sensitive to model uncertainties or external disturbances. However, the simplicity of signal generation and the independence from system observation make pacemaker control attractive in settings where sensing is difficult or unavailable.

In contrast, the mean-phase feedback control actively adjusts the control input based on real-time measurements of the mean-phase within each cluster. This enables the method to respond adaptively to the current system state and potentially achieve more flexible synchronization patterns. The proposed design method allows for the local feedback gain to be optimized cluster-wise through a convex optimization problem \eqref{eq:feedback_constraint_stability} and \eqref{eq:feedback_constraint_non-negative}, which makes it possible to reduce the number of controlled nodes or account for limitations in input placement. As a closed-loop method, this approach can also be more robust against disturbances or modeling errors. On the other hand, it requires reliable phase measurements and may suffer from implementation issues such as sensing noise or communication delays. In particular, latency in measurement or actuation can affect control performance, which is not a concern in the pacemaker method due to its open-loop nature.

Overall, the two control methods possess complementary advantages. The pacemaker offers simplicity and minimal infrastructure requirements, while the mean-phase feedback control provides adaptivity and robustness at the cost of increased sensing and computation. The choice between them depends on the available resources, the network structure, and the desired level of synchronization flexibility and reliability.

\section{Numerical Example}
\label{sec:numerical example}
In this section, we demonstrate the effectiveness of the proposed methods for stabilizing cluster synchronization and cluster phase cohesiveness through a numerical example.

\subsection{Cluster Synchronization}
\begin{figure}[tbp]
    \begin{minipage}{0.49\hsize}
        \centering
        \includegraphics[width=\hsize]{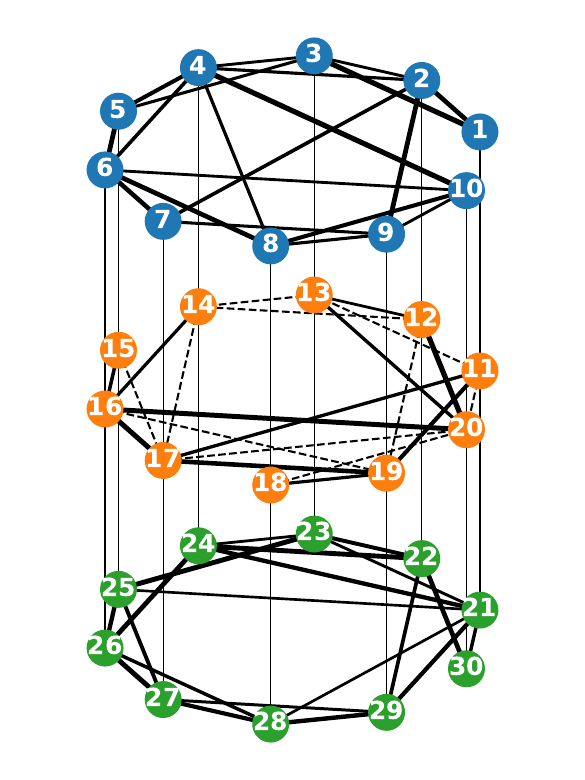}\\
        \vspace{-0.8em}
        (a) Network
    \end{minipage}
    \begin{minipage}{0.49\hsize}
        \centering
        \vspace{2em}
        \includegraphics[width=\hsize]{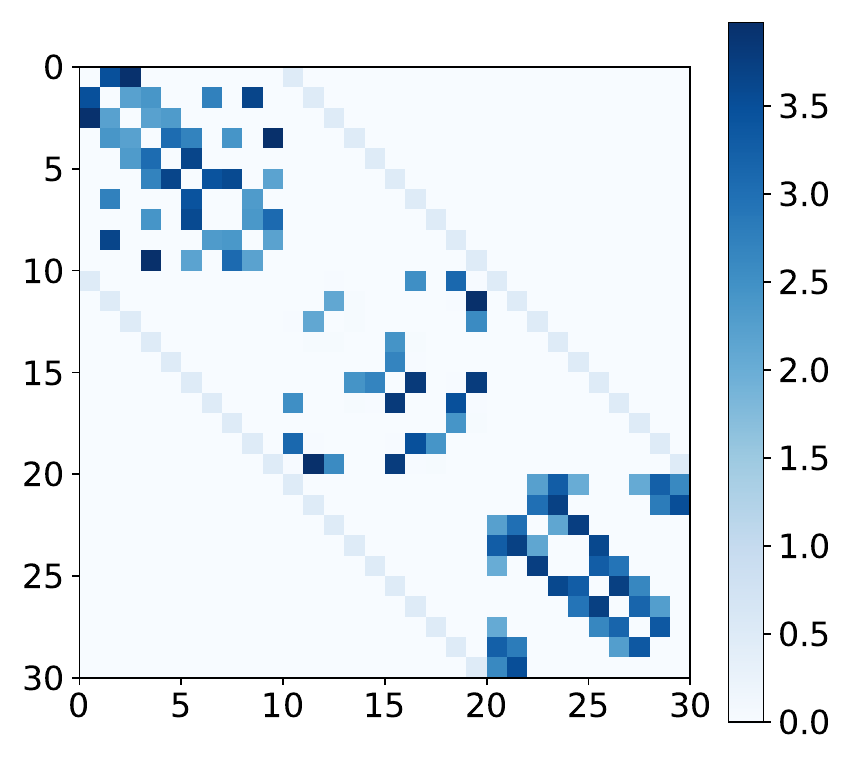}\\
        (b) Adjacency matrix
    \end{minipage}\\
    \caption{Simulation setting for cluster synchronization}
    \label{fig:simulation_network}
    \vspace{-4mm}
\end{figure}

In the first part, we examine cluster synchronization using the network depicted in Fig.~\ref{fig:simulation_network}~(a) with three clusters, where each cluster has 10 nodes.
Fig.~\ref{fig:simulation_network}~(b) shows the connection strengths among the nodes by representing the adjacency matrix in a heatmap format.
The intra-cluster connections of each cluster are generated by the Watts--Strogatz model \cite{ws-1998} and their weights are randomly chosen from the uniform distribution over the interval $[2, 4]$.
The inter-cluster connections are chosen so that Assumption~\ref{asm:cluster_sync_asm}~(ii) is satisfied.
Note that the top cluster and the bottom cluster are not connected in Fig.~\ref{fig:simulation_network} (a).
Similarly to \cite{menara-cdc2019}, we consider a scenario where, as a result of some damage to the network, half of the edges of the center cluster have been weakened in such a way that their weights are multiplied by 0.01.
In Fig.~\ref{fig:simulation_network}~(a), such edges are shown in dashed lines.

The natural frequencies associated with clusters $\Cluster_1$, $\Cluster_2$, and $\Cluster_3$ are set as $5$, $10$, and $15$, respectively.
The initial phase of each node is set randomly in $[0, 2\pi]$.
We also use $h_k = \zeros_{|\Cluster_k|}$ for each $k\in\Set{1,\,2,\,3}$.

\begin{figure}[tbp]
    \centering
    \includegraphics[width=\hsize]{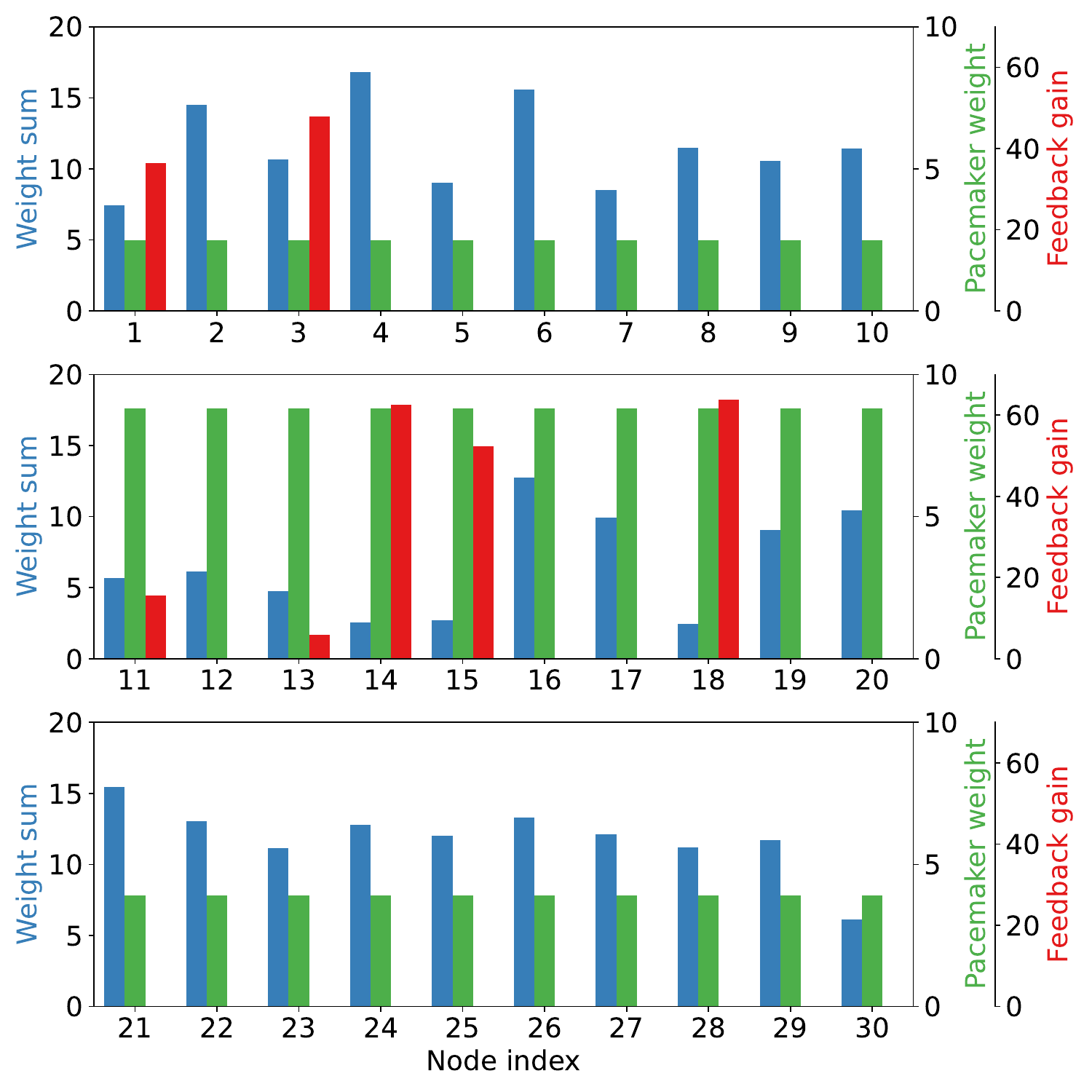}
    \vspace{-1em}
    \caption{Weight sums for the nodes, feedback gains, and pacemaker weights}
    \label{fig:simulation_parameters}
    \vspace{-7mm}
\end{figure}

In Fig.~\ref{fig:simulation_parameters}, the blue bars show the weight sums of edges that are connected to each node, the green bars show the pacemaker weights for the nodes, and the red bars show the feedback gains for the nodes.
We designed the parameters for the two proposed control methods as follows.
First, for the mean-phase feedback control, we designed the feedback gains by solving the optimization problem in Section \ref{subsec:optimization}.
For their calculation, we used the \texttt{fmincon} function in Matlab.
In the central cluster, nodes given larger values of feedback gains seem to be those having smaller sums of edge weights and fewer connections to other nodes; the feedback helps to strengthen their connections for reaching synchronization.
We emphasize that for the bottom cluster, it is not necessary to measure the mean phase, since the gains turned out to be all zero.
Second, for the pacemakers, we chose the weights based on Theorem~\ref{thm:pacemaker_gain}.
Recall that in Theorem~\ref{thm:pacemaker_gain}, the pacemaker weights are identical in each cluster.
Comparing the pacemaker weights to the feedback gains, we can see that mean-phase feedback control allows localization of the nodes to be controlled, but the feedback gains must be larger than the pacemaker weights for some nodes.

\begin{figure}[tbp]
\begin{figure}[H]
    \centering
    \includegraphics[width=0.75\hsize]{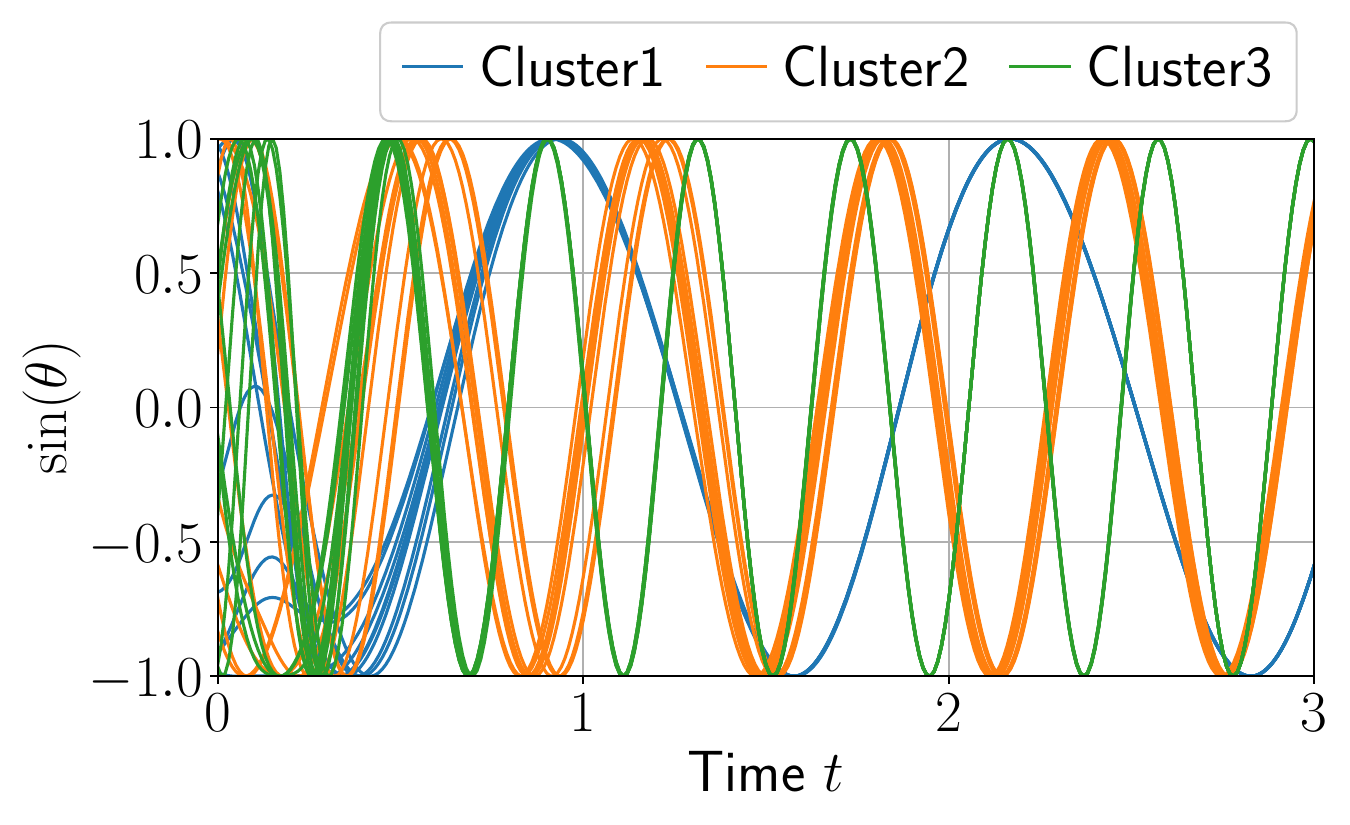}
    \vspace{-4mm}
    \caption{Time responses of phases: Uncontrolled}
    \label{fig:simulation_uncontrolled_result}
\end{figure}
\vspace{-5mm}
\begin{figure}[H]
    \centering
    \includegraphics[width=0.75\hsize]{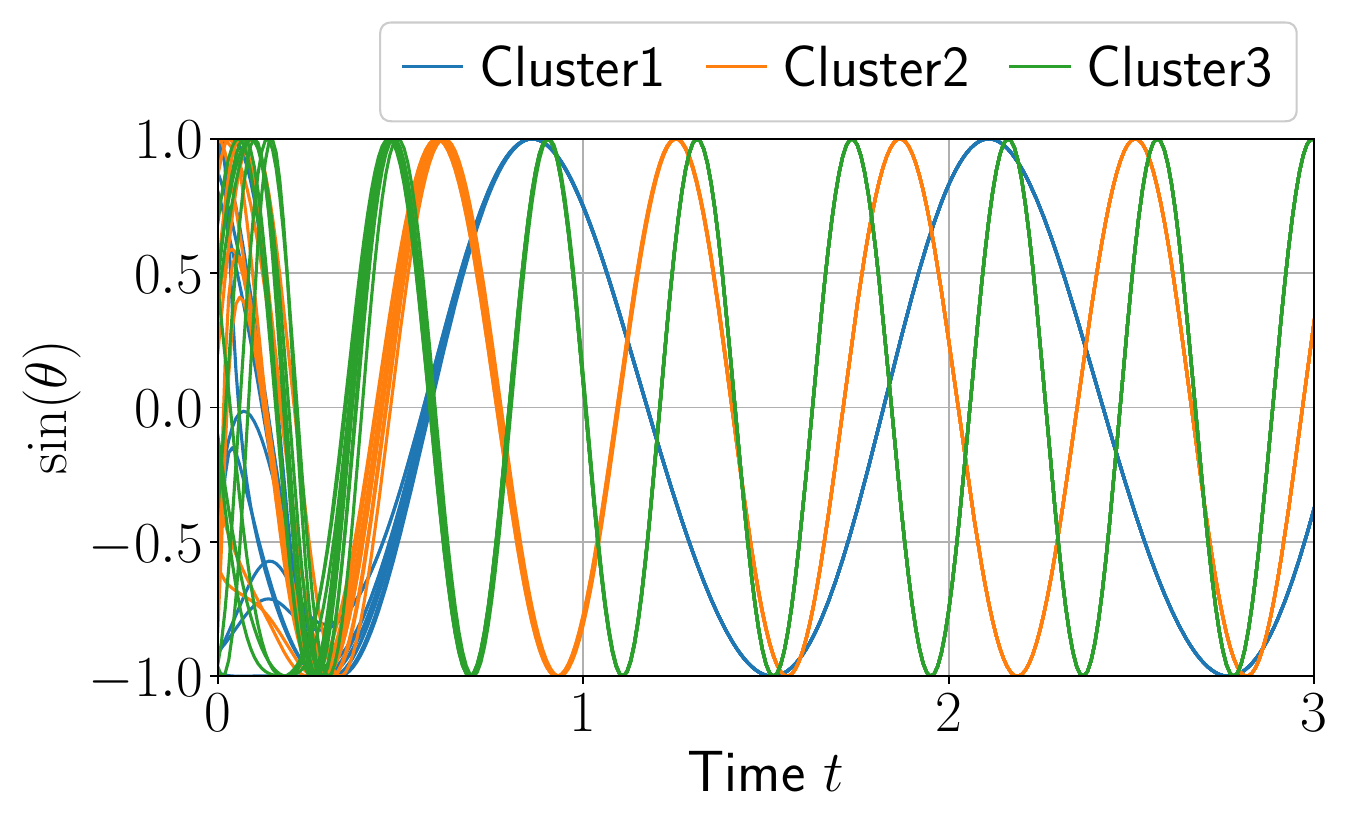}
    \vspace{-4mm}
    \caption{Time responses of phases: Mean-phase feedback control}
    \label{fig:simulation_feedback_result}
\end{figure}
\vspace{-5mm}
\begin{figure}[H]
    \centering
    \includegraphics[width=0.75\hsize]{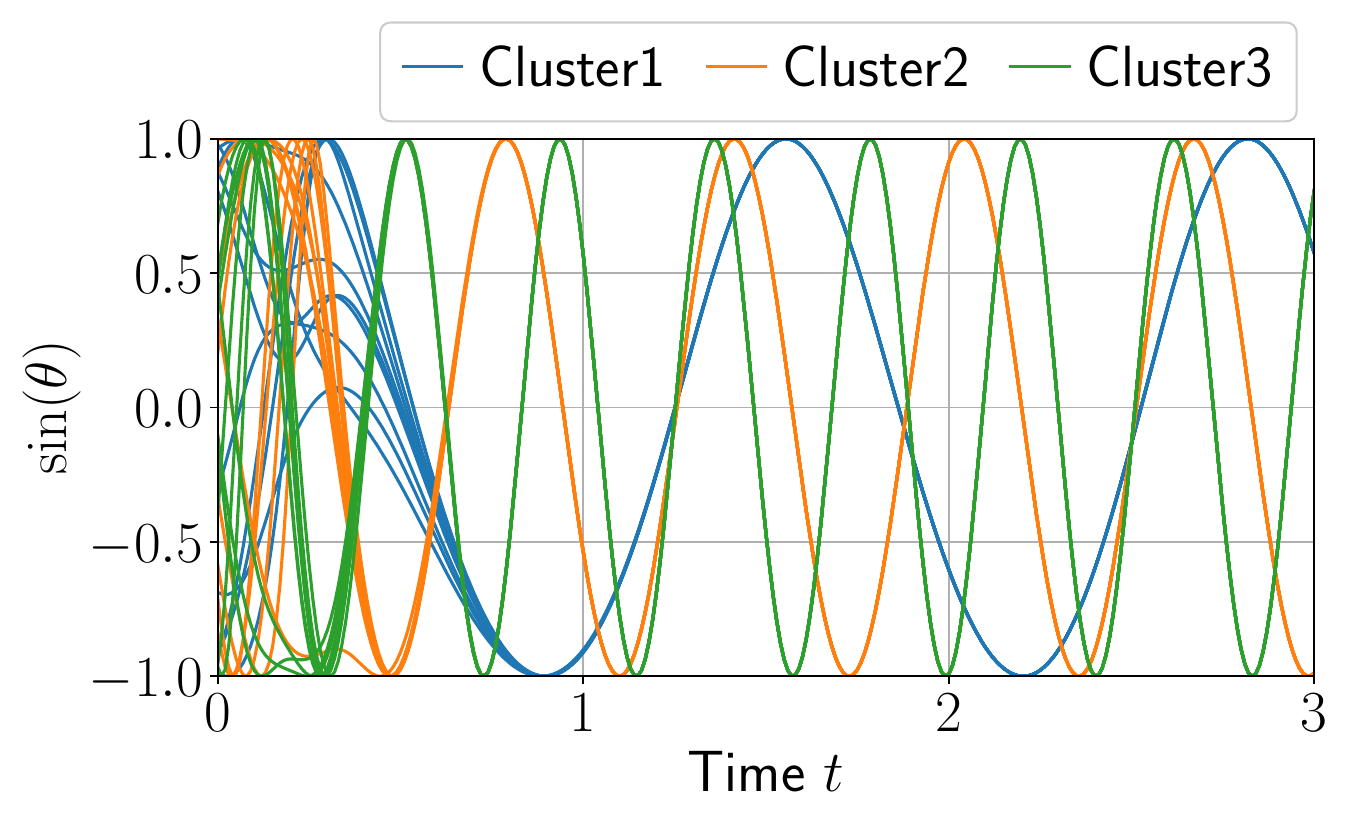}
    \vspace{-4mm}
    \caption{Time responses of phases: Pacemakers}
    \label{fig:simulation_pacemaker_result}
\end{figure}
\vspace{-6mm}
\end{figure}

Figs.~\ref{fig:simulation_uncontrolled_result}, \ref{fig:simulation_feedback_result}, and \ref{fig:simulation_pacemaker_result} show the time responses of phases of all nodes for the three cases, with no control, mean-phase feedback control, and pacemakers, respectively.
In the uncontrolled case, the central cluster (Cluster 2) reaches synchronization only approximately.
In contrast, when mean-phase feedback control and pacemakers are applied, it can be observed that the oscillators achieve cluster synchronization fairly quickly.

\subsection{Cluster Phase Cohesiveness}
\begin{figure}[tbp]
    \begin{minipage}{0.49\hsize}
        \centering
        \includegraphics[width=\hsize]{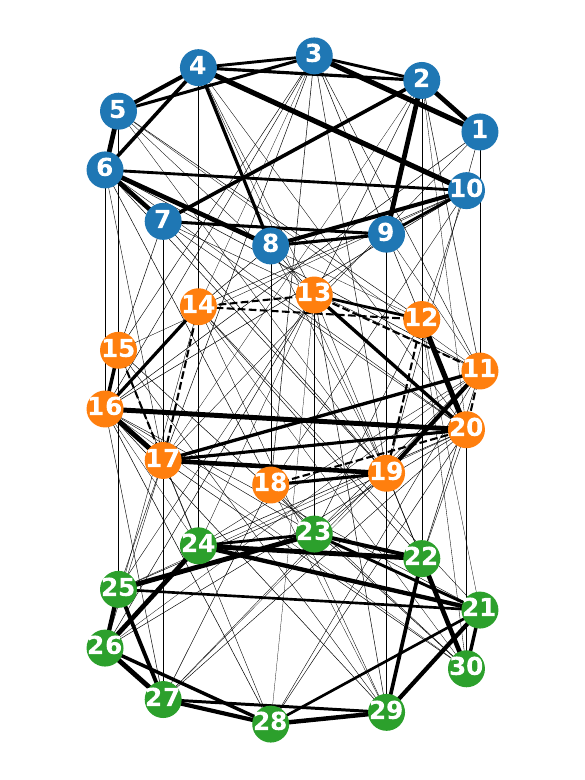}\\
        \vspace{-0.8em}
        (a) Network
    \end{minipage}
    \begin{minipage}{0.49\hsize}
        \centering
        \vspace{2em}
        \includegraphics[width=\hsize]{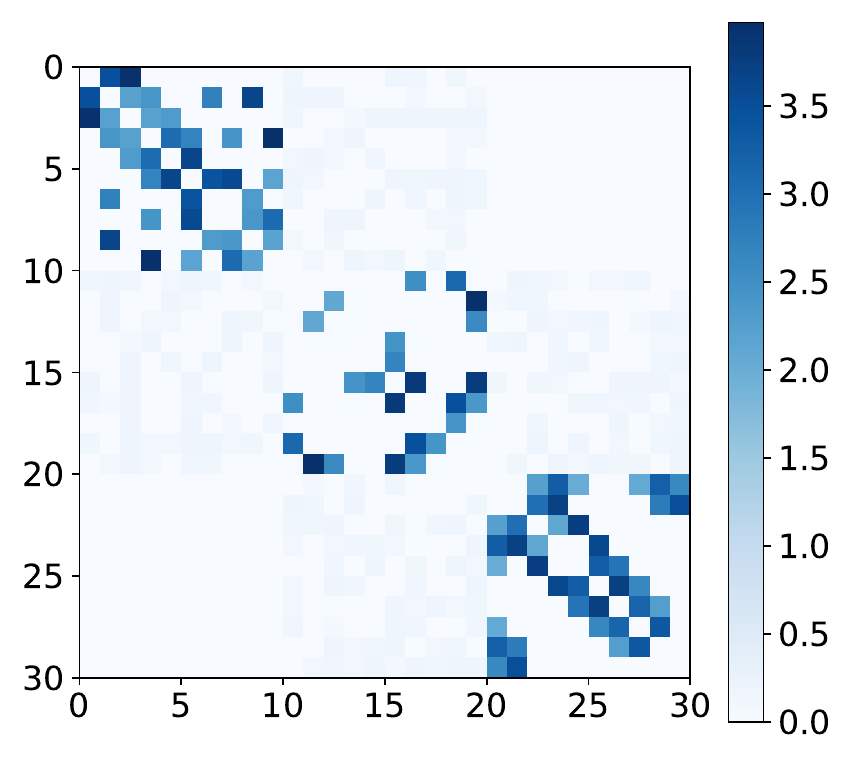}\\
        (b) Adjacency matrix
    \end{minipage}\\
    \centering
    \caption{Simulation setting for cluster phase cohesiveness}
    \label{fig:network_CPC}
\end{figure}

Next, we study cluster phase cohesiveness using the network shown in Fig.~\ref{fig:network_CPC}~(a).
This network has a structure similar to the one in Fig.~\ref{fig:simulation_network}~(a), but within each cluster, there are more edges and the natural frequencies are non-uniform.
Specifically, there are three clusters with 10 nodes each.
The intra-cluster connections of each cluster are the same as those in Fig.~\ref{fig:simulation_network}~(a).
On the other hand, the inter-cluster connections are randomly chosen with probability 0.5; moreover, their weights are randomly chosen from the uniform distribution over the interval [0.1, 0.2].
Fig.~\ref{fig:network_CPC}~(b) shows the connection strengths among nodes. 
The natural frequencies of the nodes in cluster $k$ are randomly chosen from the normal distribution $\mathcal{N}(\mu_k, 1)$, where $\mu_1 = 5$, $\mu_2 = 10$, $\mu_3 = 15$.

\begin{figure}[tbp]
\begin{figure}[H]
    \centering
    \includegraphics[width=0.75\hsize]{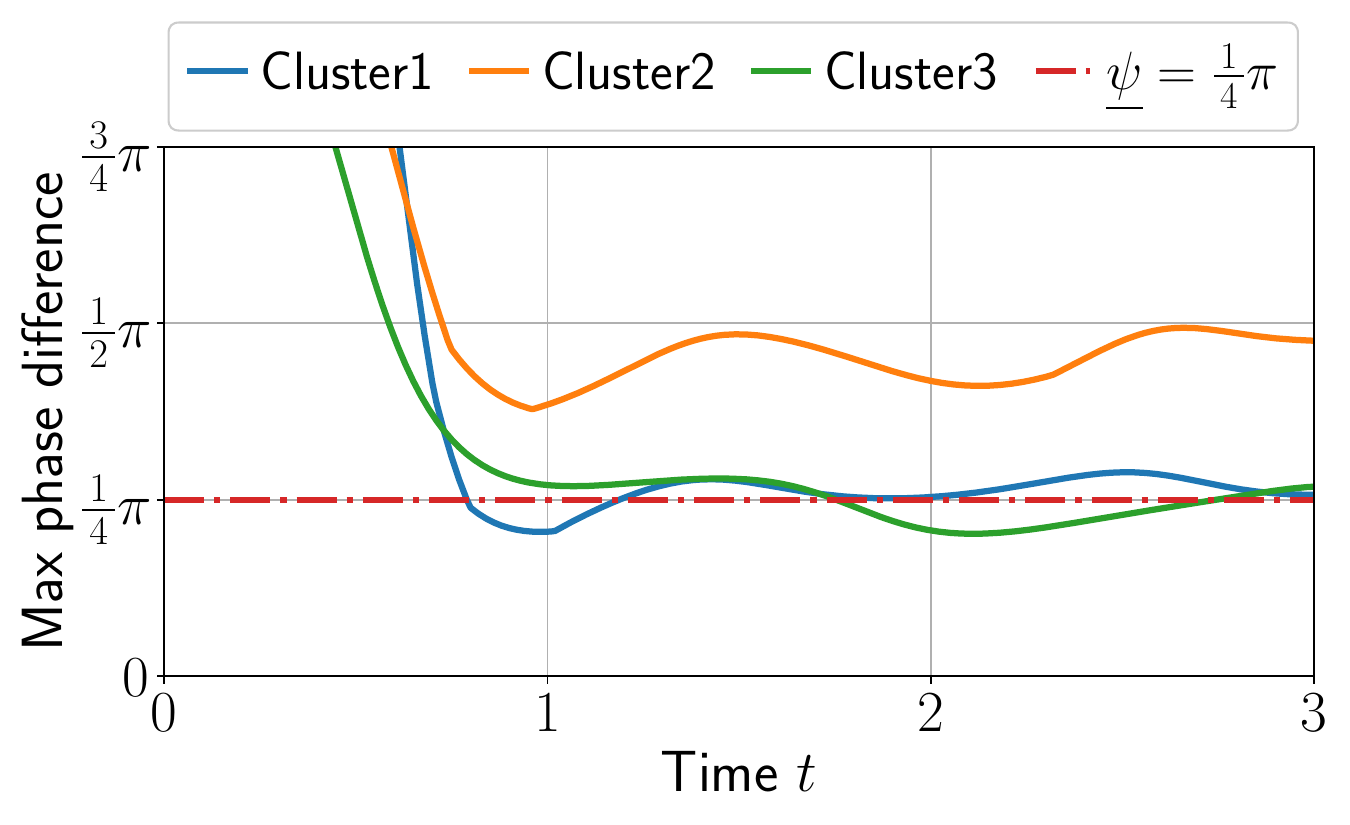}
    \vspace{-4mm}
    \caption{Time responses of maximum phase differences: Uncontrolled}
    \label{fig:result_CPC_uncontrolled}
\end{figure}
\vspace{-5mm}
\begin{figure}[H]
    \centering
    \includegraphics[width=0.75\hsize]{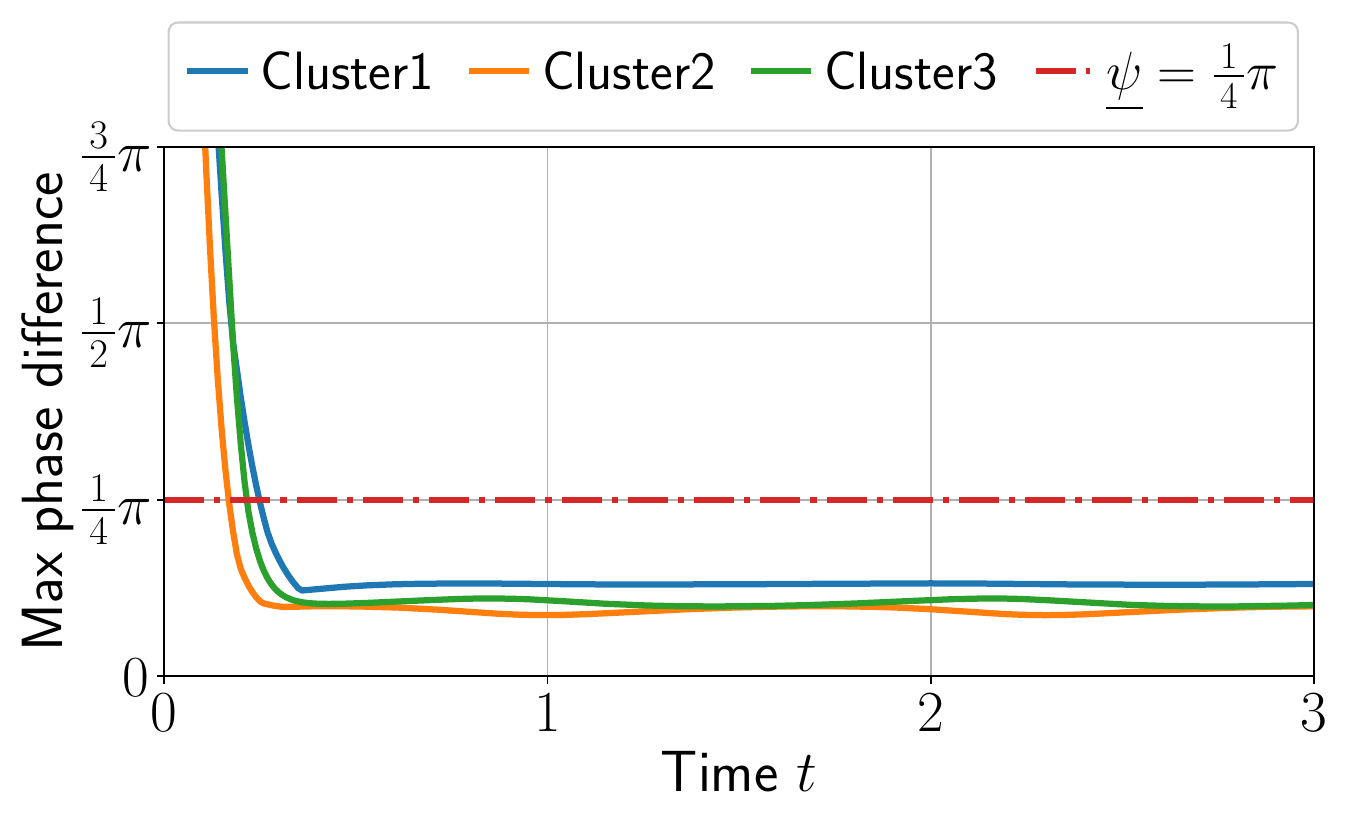}
    \vspace{-4mm}
    \caption{Time responses of maximum phase differences: Mean-phase feedback control}
    \label{fig:result_CPC_feedback}
\end{figure}
\vspace{-5mm}
\begin{figure}[H]
    \centering
    \includegraphics[width=0.75\hsize]{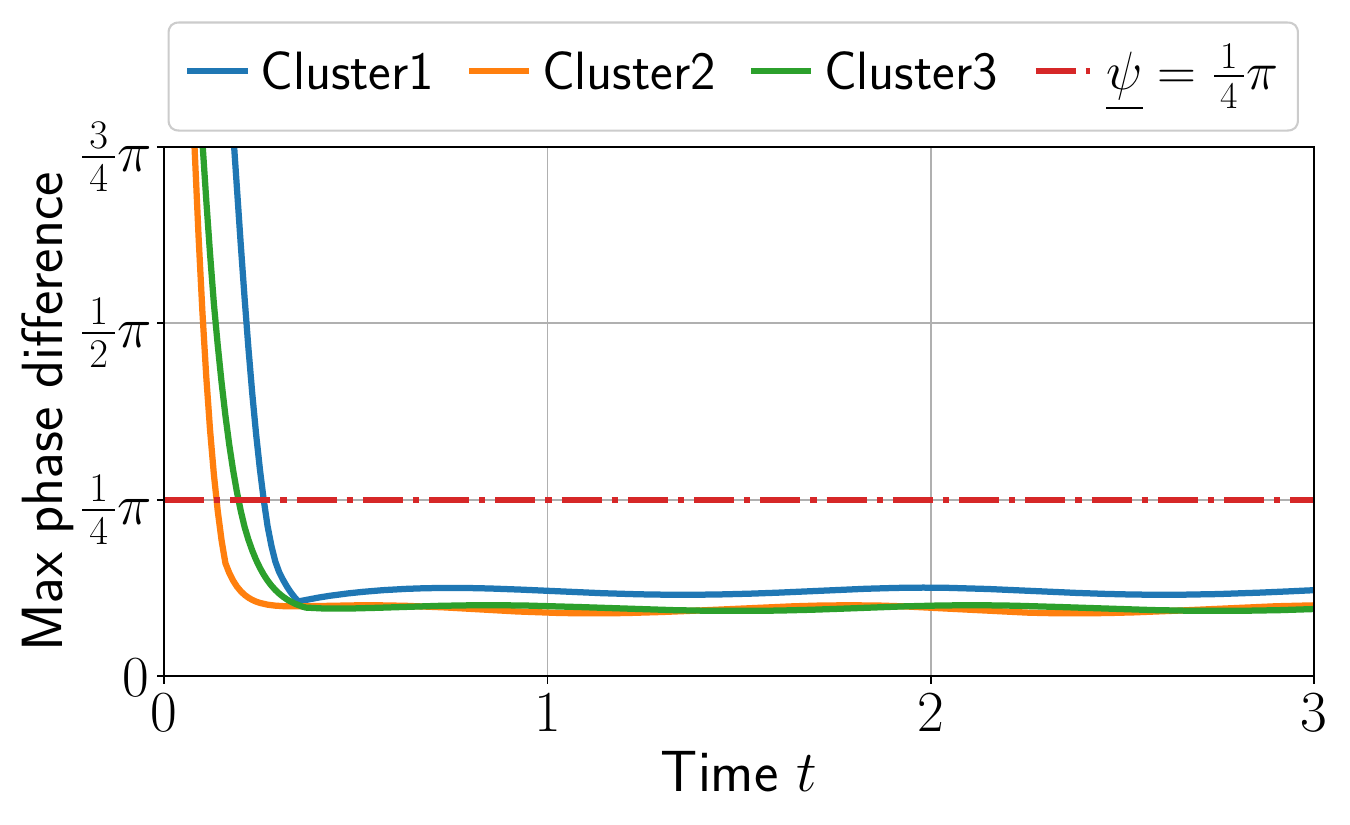}
    \vspace{-4mm}
    \caption{Time responses of maximum phase differences: Pacemakers}
    \label{fig:result_CPC_pacemaker}
\end{figure}
\end{figure}

Figs.~\ref{fig:result_CPC_uncontrolled}, \ref{fig:result_CPC_feedback}, and \ref{fig:result_CPC_pacemaker} show the time responses of the maximum phase difference for the three cases, with no control, mean-phase feedback control, and pacemakers, respectively.
Feedback gains and pacemaker weights of each cluster are chosen based on \eqref{eq:cpc_feedback_main_result} and \eqref{eq:cpc_pacemaker_main_result}, respectively, with $\underline{\psi} = \pi / 4$.
In the uncontrolled case, the phases of each cluster clearly do not reach the desired level of phase cohesiveness.
However, when mean-phase feedback control and pacemakers are applied, the phase difference in each cluster become small and reach the cohesive level of $\pi/4$ as specified in a short time period.

\section{Conclusion}
\label{sec:conclusion}
In this paper, we have studied the problem of realizing cluster synchronization of Kuramoto oscillators by introducing control signals via two approaches: pacemakers and mean-phase feedback.
We have developed conditions for pacemaker weights and feedback gains to achieve cluster synchronization as well as cluster phase cohesiveness.
We have also proposed the design method of feedback gains through convex optimization.
In future research, we will study less conservative conditions to reduce the size of the gains in these control approaches.

\smallskip\noindent
\textsl{Acknowledgment}:~The authors thank Mr Mitsuaki Matsubara for the helpful discussions on the results of the paper.

\end{document}